%% file: sample-sigconf.tex
\documentclass[sigconf]{acmart}
\usepackage{url}
\usepackage{enumitem}
\usepackage{listings}
\usepackage{color}
\usepackage{fancybox}
\usepackage{algorithm}
\usepackage{algorithmic}
\usepackage{multirow}
\usepackage{subfigure}
\usepackage{hhline}
\usepackage{amsthm} 
\usepackage{colortbl}
\usepackage{geometry}
\usepackage{makecell,rotating}
\usepackage{graphicx}

\setcopyright{none}
\renewcommand\footnotetextcopyrightpermission[1]{} 

\begin{document}

\copyrightyear{2022}
\acmYear{2022}
\setcopyright{acmcopyright}\acmConference[ASE '22]{37th IEEE/ACM International
Conference on Automated Software Engineering}{October 10--14, 2022}{Rochester,
MI, USA}
\acmBooktitle{37th IEEE/ACM International Conference on Automated Software
Engineering (ASE '22), October 10--14, 2022, Rochester, MI, USA}
\acmPrice{15.00}
\acmDOI{10.1145/3551349.3560420}
\acmISBN{978-1-4503-9475-8/22/10}

\shadowsize1.5pt
\definecolor{mygray}{gray}{.9}
\definecolor{mycyan}{cmyk}{.3,0,0,0}
\definecolor{mauve}{rgb}{0.58,0,0.82}
\definecolor{dkgreen}{rgb}{0,0.6, 0}
\definecolor{gray}{rgb}{0.5,0.5,0.5}
\lstset{
	language=Java,
	aboveskip=0mm,
	belowskip=1mm,
	showstringspaces=true,
	columns=flexible,
	basicstyle={\scriptsize\ttfamily },
	frame=tb,
	boxpos = c,
	numberstyle={\footnotesize},
	alsoletter=-,
	morekeywords={Preference, Bundle, Object, this, manifest, activity, application, intent-filter, action, category, data, @Override},
	keywordstyle=\color{blue},
	commentstyle=\color{dkgreen},
	stringstyle=\color{mauve}\ttfamily,
	breaklines=true,
	breakatwhitespace=true,
	showstringspaces=true,
	tabsize=2,
	escapeinside=``
}
\setenumerate[1]{itemsep=0pt,partopsep=0pt,parsep=\parskip,topsep=1pt}
\setitemize[1]{itemsep=0pt,partopsep=0pt,parsep=\parskip,topsep=1pt}
\setdescription{itemsep=0pt,partopsep=0pt,parsep=\parskip,topsep=1pt}


\renewcommand{\algorithmicrequire}{\textbf{Input:}}
\renewcommand{\algorithmicensure}{\textbf{Output:}}
\newcommand{\tabincell}[2]{\begin{tabular}{@{}#1@{}}#2\end{tabular}}

\newif\ifdraft\drafttrue
\ifdraft
\newcommand{\revise}[1]{\color{black}{#1}\color{black}}
\else
\newcommand{\revise}[1]{}
\fi

\title{A Comprehensive Evaluation of Android ICC Resolution Techniques}

\author{Jiwei Yan}
\affiliation{%
	\department{Tech. Center of Software Engineering}
	\institution{Institute of Software, CAS, China}
	\institution{Univ. of Chinese Academy of Sciences}
	\city{Beijing}
	\country{China}
}
\email{yanjw@ios.ac.cn}

\author{Shixin Zhang}
\affiliation{%
	\institution{School of Software Engineering\\Beijing Jiaotong University}
	\city{Beijing}
	\country{China}
}
\email{zhangsx@bjtu.edu.cn}

\author{Yepang Liu}
\authornote{Yepang Liu is affiliated with the Research Institute of Trustworthy Autonomous Systems, Guangdong Provincial Key Laboratory of Brain-inspired Intelligent Computation, and Department of Computer Science and Engineering of SUSTech.}
\affiliation{%
	\institution{Dept. of Computer Science and Engr. Southern University of Sci. and Tech.}
	\city{Shenzhen}
	\country{China}
}
\email{liuyp1@sustech.edu.cn}

\author{Xi Deng}
\affiliation{%
	\department{State Key Lab. of Computer Science}
	\institution{Institute of Software, CAS, China }
	\institution{Univ. of Chinese Academy of Sciences}
	\city{Beijing}
	\country{China}
}
\email{dengxi@ios.ac.cn}

\author{Jun Yan}
\authornote{Corresponding Authors.}
\affiliation{%
	\department{State Key Lab. of Computer Science}
	\institution{Institute of Software, CAS, China }
	\institution{Univ. of Chinese Academy of Sciences}
	\city{Beijing}
	\country{China}
}
\email{yanjun@ios.ac.cn}

\author{Jian Zhang}
\authornotemark[2]
\affiliation{%
	\department{State Key Lab. of Computer Science}
	\institution{Institute of Software, CAS, China }
	\institution{Univ. of Chinese Academy of Sciences}
	\city{Beijing}
	\country{China}
}
\email{zj@ios.ac.cn}
\begin{abstract}
Inter-component communication (ICC) is a widely used mechanism in mobile apps, which enables message-based control flow transferring and data passing between Android components.
Effective ICC resolution requires precisely identifying entry points, analyzing data values of ICC fields, modeling related framework APIs, etc.
Due to various control-flow- and data-flow-related characteristics involved and the lack of oracles for real-world apps, the comprehensive evaluation of ICC resolution techniques is challenging.

To fill this gap, we collect multiple-type benchmark suites with 4,104 apps, covering hand-made apps, open-source, and commercial ones.
Considering their differences, various evaluation metrics, e.g., number count, graph structure, and reliable oracle based metrics, are adopted on-demand. 
As the oracle for real-world apps is unavailable, we design a dynamic analysis approach to extract the real ICC links triggered during GUI exploration.
By auditing the code implementations, we carefully check the extracted ICCs and confirm 1,680 ones to form a reliable oracle set, in which each ICC is labeled with 25 code characteristic tags. 
The evaluation performed on six state-of-the-art ICC resolution tools shows that 1) the completeness of static ICC resolution results on real-world apps is not satisfactory, as up to 38\%-85\% ICCs are missed by tools;
2) many wrongly reported ICCs are sent from or received by only a few components and the graph structure information can help the identification;
3) the efficiency of fundamental tools, like ICC resolution ones, should be optimized in both engineering and research aspects.
By investigating both the missed and wrongly reported ICCs, we discuss the strengths of different tools for users and summarize eight common FN/FP patterns in ICC resolution for tool developers.

\end{abstract}

\keywords{Android, Inter-Component Communication (ICC), Transition Graph}

\maketitle

\input{content/introduction}
\input{content/background}
\input{content/tooloverview}

\input{content/dataCollection}

\input{content/evaluation}
\input{content/patterns}

\input{content/discuss}

\input{content/conclusion}

\bibliographystyle{ACM-Reference-Format}
\bibliography{bib/ref}

\end{document}

%% file: content/introduction.tex
\section{Introduction}
   
\revise{Android programs are composed of four types of basic components, which are provided to interact with users, perform background tasks, etc.
Each component is a single module and components communicate with each other through the Inter-component communication (ICC) mechanism. 
To figure out the control and data flows between the source and target components, users can use ICC resolution tools to extract ICC-related information. }
The most widely used ICC resolution tools are \textit{Epicc}~\cite{DBLP:conf/uss/OcteauMJBBKT13} and \textit{IC3}~\cite{Octeau15ICSE}. 
They model the ICC-related framework APIs and perform a data-flow analysis to resolve the ICC field values, whose results can be used to construct the component/activity transition graph (CTG/ATG).
Some works~\cite{DBLP:conf/icse/ChenFCSLLX19, chen2019storydistiller} use the constructed transition graph to help the program behavior understanding, 
and others~\cite{DBLP:conf/oopsla/AzimN13,DBLP:conf/icse/JabbarvandLM19, DBLP:conf/kbse/FanSCMLXP18} use it to help automatic test generation.
Besides, there are various ICC-related vulnerabilities that have attracted the attention of researchers, 
including inter-component privacy leak \cite{Li15ICSE, DBLP:conf/iscisc/BohluliS18,DBLP:journals/compsec/ZhangTD21}, 
permission leak \cite{DBLP:journals/tse/BagheriSGM15, DBLP:conf/icse/SadeghiJGBM18, DBLP:journals/tr/ZhangTDZ21}, 
and inter-app collusion \cite{DBLP:conf/ccs/BosuLYW17, DBLP:conf/icse/LeeBSSZM17,DBLP:journals/tmc/ElishCBYR20,DBLP:journals/fgcs/BhandariHLZGR20}. 
With wide usage in various scenarios, both the soundness and completeness of ICC resolution results have great impact on its applications.


In Android, an ICC message is represented as an \texttt{Intent} \cite{Intent} object, which contains a set of Intent fields.
To obtain the source components of ICCs, we need to find the control flows from entry point methods to the \texttt{Intent} object sending statements, where the entry method may be user-customized ones that are difficult to be identified.  
And to find out the target component, the values of carried Intent fields should be carefully analyzed.
As many code characteristics are involved when identifying the source and target of ICC messages, resolving ICCs with high precision is a challenging task.
During analysis, the imprecision in any step, i.e., while handling any code characteristic, may lead to either false positives (FPs) or false negatives (FNs) in the final results.
Moreover, these FNs or FPs may be propagated upwards as ICC resolution usually works as a fundamental module, e.g., more than half of FNs in \textit{LogExtractor}~\cite{DBLP:journals/di/ChengSGG21} are ICC-related as it invokes the ICC resolution tool \textit{IC3}~\cite{Octeau15ICSE}.
Actually, for both the users and developers of ICC resolution tools, it is hard to know whether the reported ICCs are trustworthy or not and the root causes of precision loss. 
Therefore, to figure out that, a comprehensive evaluation focusing on Android ICC resolution techniques is required.


There are several off-the-shelf benchmarks~\cite{ICC-Bench, DroidBench} that can be reused for ICC-related evaluation. They are designed by researchers who want to measure tools' effectiveness when encountering ICC code snippets.
Although widely adopted, it is questionable whether these hand-made apps could represent complex real-world ones, for they are designed only with a few code characteristics. 
For a more practical evaluation based on real-world codes, there are two challenges.
Lacking proper metrics is the \textbf{first challenge}.
When evaluating apps without available ground truths, many works~\cite{DBLP:conf/uss/OcteauMJBBKT13, Octeau15ICSE, DBLP:conf/ccs/BosuLYW17} measure and compare the number of resolved ICC links instead.
The number-based comparison is effective only when the tools under evaluation rarely report nonexistent ICCs, i.e., FPs.
However, according to the further experimental results on hand-made apps, FPs exist for most ICC resolution tools, which makes the number-based comparison less convincing.
\textbf{Another challenge} is the lack of high-quality benchmark suites.
A high-quality benchmark suite requires both the representative test inputs, i.e., Android apps, and available test oracle, i.e., ground truth ICCs.
To figure out the different behaviors of tools when resolving ICCs with various code characteristics, 
both the ICC-related code snippets and the involved code characteristics should be identified and labeled for each ICC in the test oracle. 
Such information can also help developers to find real-world instances for each unhandled characteristic, and give directions for tool updating. However, there are no such characteristics-labeled benchmark suites up to now.



In this paper, we focus on the comprehensive evaluation of widely used ICC resolution tools and pick six state-of-the-art ones as the evaluation subjects.
We collect multiple-type benchmark suites with 4,104 apps, including hand-made app sets, large-scale real-world ones, as well as a compact but representative dataset with reliable oracles, with which we can observe tools' performance on different app sets.
For different benchmarks, we adopt different evaluation metrics, in which the \textit{number-based} metrics have weak credibility but strong versatility, so they fit all the benchmarks; 
the \textit{graph-based} metrics require that ICCs are actually designed for real mobile users, so they are suitable for real-world apps but not hand-made ones;
and the \textit{oracle-based} metrics should be applied on datasets with available oracles and code characteristic labels.

For hand-made apps, we can easily label their ICCs as well as the ICC-related code characteristics.
As oracles for real-world apps are unavailable, we design a dynamic analysis approach to collect as many real ICC links as possible, because the dynamically triggered ICCs are not limited by the complexity of static code characteristics.
First, we adopt the GUI exploration approach to trigger ICCs, which covers 58.9\% app components.
By monitoring the execution traces of apps, we propose a specific ICC extraction approach considering the ICC launching characteristics. 
And to ensure the reliability of collected ICCs, we combine automatic result filters and careful manual code auditing.
Finally, we successfully map 1,680 ICCs to corresponding code snippets and label the involved code characteristics, which form a reliable benchmark with ICCs extracted from real-world apps. The apps, ICC oracles and their code characteristic tags are all publicly available \href{https://iccviewer.ldby.site/ICCViewer/}{\underline{here} } \cite{ICCViewer,ICC-Resolution-Evaluation}.

%
Through the evaluation, we have the following observations.
First, tools behave inconsistently on multiple benchmarks, which reflects the necessity to construct reliable oracles on real-world apps. 
Especially, the completeness of ICC resolution results on real-world apps is not satisfactory. Up to 38\%-85\% ICCs are missed by tools for their inadequate analysis of specific code characteristics.
Second, many fake ICCs are sent from or received by only a few components and number-based metrics could not identify them.
With the help of graph-based metrics, we can quickly identify a set of wrongly reported ICCs, which are caused by conservative analysis or the transitivity of imprecision.
Besides, most tools suffer from the inefficiency problem when working on complex real-world apps. Users don't know when the analysis will finish and have to terminate it with no output.
Finally, based on the evaluation results, we recommend typical scenarios of tool usage and summarize eight common FN/FP patterns in ICC resolution.


\textbf{Contributions.} The contributions of this work are threefold:
\begin{itemize}[leftmargin=10pt]
\item We construct multiple-type benchmark suites for ICC resolution, which contain both hand-made apps designed with specific characteristics and real-world apps with complex ICC implementation, and propose a dynamic ICC extraction approach to obtain characteristic-labeled oracles for representative apps.
\item We propose a unified ICC resolution comparison framework and design specific metrics for multiple-type benchmark suites.
\item We carry out in-depth evaluations on six popular and state-of-the-art ICC resolution tools, clarify the strengths and weaknesses of each tool, summarize the root causes that lead to precision loss, and discuss the directions for further improvement.
\end{itemize}



%% file: content/background.tex
\begin{table*}[!htbp]
	\centering
	\footnotesize
	\setlength{\abovecaptionskip}{5pt}
	\setlength{\belowcaptionskip}{-10pt}
	\caption{Overview of the ICC Resolution Tools (\checkmark: True, X: False, -: Unknown, *: To be Discussed)}\label{tab:tools}
	\begin{tabular}{|c|c|c|c|l|l|l|c|c|c|c|c|c|c|c|c|c|c|}
	\hline  
	\textbf{Tool } &\textbf{\tabincell{c}{Last\\Update}} &\textbf{\tabincell{c}{Apk\\input}} &\textbf{\tabincell{c}{Graph\\Output}} &\textbf{Functionality} 
	&\textbf{\tabincell{c}{Base Tool/\\Framework}} & \textbf{Approach} &\textbf{\tabincell{c}{Sensitivity\\F/ C/ I/ O}}  
	&\textbf{\tabincell{c}{Component\\Act/NAct/Fr}} & \textbf{\tabincell{c}{Extra Data}} 	&\textbf{\tabincell{c}{StrOp}}  \\\hline

	\textbf{A$^3$E} \cite{a3e} 	&2016-09 &\checkmark &ATG*&ATG Construction 	 &SCanDroid/ Wala   &Taint Analysis&-   &\checkmark/ \checkmark*/ X &X &- \\ \hline
	\textbf{IC3} \cite{IC3} 	&2015-09 &\checkmark &X	&ICC Resolution 		 &Epicc/ Soot   &IDE Analysis&\checkmark/ \checkmark*/ \checkmark/ X  &\checkmark/ \checkmark/ X &\checkmark &\checkmark  \\ \hline
\textbf{IC3$_{Dial}$} \cite{IC3-DIALDroid} &2020-02 &\checkmark &X &ICC Resolution 		 &IC3/ Soot   &IDE Analysis&\checkmark/ \checkmark*/ \checkmark/ X  &\checkmark/ \checkmark/ X &\checkmark &\checkmark \\ \hline
	\textbf{Gator} \cite{GATOR} 	&2019-05 &\checkmark &ATG &ATG Construction	 &/Soot   &IDE Analysis* &X/ X/ -/ - &\checkmark/ X / X &\revise{X} &X \\ \hline
	\textbf{StoryD} \cite{StoryDistiller}	&2022-03 &\checkmark &ATG* &Storyboard Generation 	 &IC3/ Soot   &IDE Analysis* &(\checkmark/ \checkmark/ \checkmark/ X)*&\checkmark/ \checkmark*/ \checkmark &X &\checkmark \\ \hline
	\textbf{ICCBot} \cite{ICCBot}	&2022-04 &\checkmark &CTG &ICC Resolution 	 &/Soot   &\tabincell{c}{Slice, Summary}&\checkmark/ \checkmark/ \checkmark/ \checkmark  &\checkmark/ \checkmark/ \checkmark &\checkmark &\checkmark \\ \hline
  \end{tabular} 
  \vspace{-0.2cm}
\end{table*}

\section{ICC Resolution: An Overview}
This section gives an overview of the ICC mechanism, the state-of-the-art ICC resolution tools and the widely used metrics for ICC evaluation.

\subsection{Overview of ICC Mechanism} \label{overview}

\begin{figure} [htbp!]
	\setlength{\abovecaptionskip}{5pt}
	\setlength{\belowcaptionskip}{-10pt}
	\includegraphics[width=0.48\textwidth]{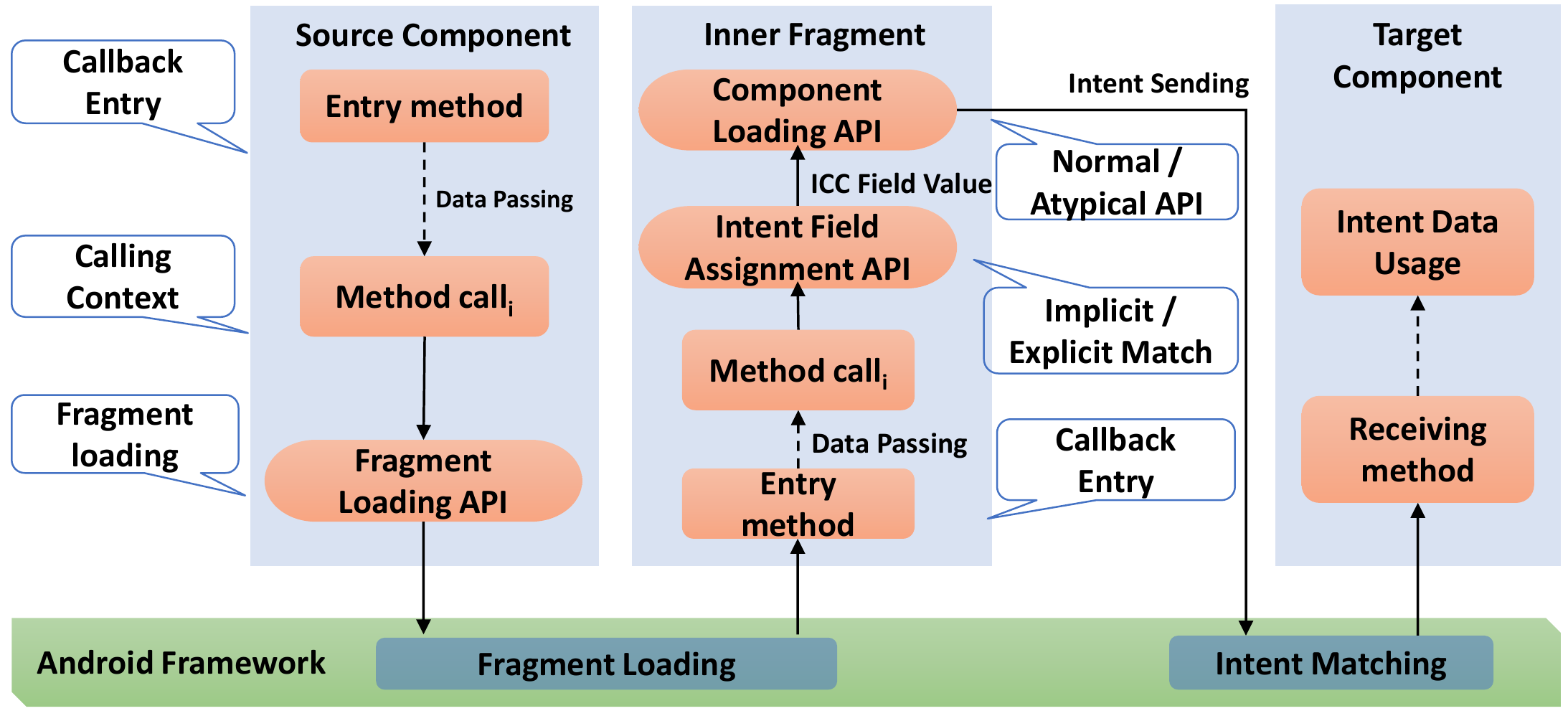}
	\caption{Overview of ICC Sending Process}
	\label{fig:Example} 
  \end{figure}

There are four basic components~\cite{component} in Android apps, including \texttt{Activity}, \texttt{Service}, \texttt{BroadcastReceiver}, and \texttt{ContentProvider}.
For the convenience of communication among app components, the Android system provides the ICC mechanism. 
An app component can create an \texttt{Intent} object and send it to the Android system.
Along with the Intent, both the basic ICC fields, e.g., \texttt{action} and \texttt{category}, and the user-customized extra data fields will be delivered.
The system resolves the value of these fileds to infer the target components.
Fig. ~\ref{fig:Example} displays the overview of the ICC sending process with five commonly used characteristics in ICC resolution.

\textbf{Callback Entry.} 
Unlike Java programs, there is no main method in an Android applications.
Instead, each component has a set of entry methods, including lifecycle methods (e.g., \texttt{onCreate()}) and callbacks (e.g., \texttt{onClick()}).
Both of them are invoked by the Android system in response to GUI or system events.
\revise{The callback recognition is challenging, for callbacks could be registered in Android framework classes or libraries (implicit callback), registered in code of application package (dynamic callback), or declared in the XML files (static callback).
And the ICC invocation may be hidden behind a series of complex callback triggering.}


\textbf{Calling Context}.
The data extraction of the Intent object fields is essential to ICC resolution. 
Considering both the Intent object itself and the data of Intent fields can be passed among function calls, a context-sensitive inter-procedural analysis should be performed.

\textbf{Fragment}.
\revise{Fragment is a dynamically loaded building block of an app's user interface, which }is hosted by an activity or another fragment~\cite{fragment}.
In Fig. ~\ref{fig:Example}, the source component first loads an inner fragment \texttt{f}, i.e., invokes its callback \texttt{c}.
Then, \texttt{f} sends an \texttt{Intent} out in one of its methods, which is reachable from \texttt{c}.
In this case, without fragment modeling, the entry point tracking analysis may terminate at the lifecycle method of a fragment, but miss the actual entry method in its host activity.


\textbf{Implicit Match}. 
\revise{There are two types of Intent.
For explicit Intent, their target components can be obtained by analyzing the value of the class or component name related fields.
For implicit Intent, the values of fields, e.g., action and category, that are related to implicit matching will determine the destination class.}

\textbf{Atypical ICC}.
For each ICC, it will be delivered to the Android system through specific API, i.e., the exit point.
\revise{Besides normally used ICC sending APIs, like \texttt{startActivity()}, there are many atypical ICC-related APIs \cite{DBLP:conf/icse/SamhiBBK21} in the Android framework, which also work as exit points although the official Android documentation does not specifically discuss them, e.g., \texttt{sendIntent()}. }
These atypical ICCs can also establish ICC links and should be concerned during ICC resolution.




%% file: content/tooloverview.tex
\subsection{Existing Tools and Application Scenarios}
Researchers have proposed many works that apply the ICC resolution results.
One of the most popular application scenarios is security property checking, especially privacy leak detection.
In the beginning, the intra-component leak detection~\cite{Arzt14PLDI, Leakminer, SCANDAL} is concerned.
Considering that many sensitive data are passed by ICC messages, researchers extend the approaches to support inter-component leak detection, including \textit{IccTA}~\cite{Li15ICSE}, \textit{Amandroid}~\cite{DBLP:conf/ccs/WeiROR14} and \textit{DroidSafe}~\cite{DBLP:conf/ndss/GordonKPGNR15}.
Besides privacy leak~\cite{ DBLP:conf/iscisc/BohluliS18,DBLP:journals/compsec/ZhangTD21}, ICC resolution also relates to permission leak~\cite{DBLP:journals/tse/BagheriSGM15, DBLP:conf/icse/SadeghiJGBM18, DBLP:journals/tr/ZhangTDZ21},
inter-app collusion~\cite{DBLP:conf/ccs/LuLWLJ12,DBLP:conf/ndss/ZhangY14, DBLP:conf/icse/LeeBSSZM17,DBLP:journals/compsec/BhandariJJLZGMC17, DBLP:conf/sigsoft/TangSWLZ020, DBLP:conf/ccs/BosuLYW17, DBLP:journals/tmc/ElishCBYR20,DBLP:journals/fgcs/BhandariHLZGR20}, etc.
Another typical scenario are GUI testing, e.g., using the constructed transition model to guide the target-directed test generation~\cite{DBLP:conf/kbse/LaiR19, DBLP:conf/icse/JabbarvandLM19, DBLP:conf/kbse/FanSCMLXP18}, and generating the storyboard of apps~\cite{DBLP:conf/icse/ChenFCSLLX19,chen2019storydistiller}.

With the wide usage of ICC, we start a systemic investigation around ICC resolution from two well-known works, \textit{IC3} \cite{Octeau15ICSE,DBLP:journals/tse/OcteauLJM16} and \textit{Epicc} \cite{DBLP:conf/uss/OcteauMJBBKT13}. Among all their citations, we first filter the works without mentioning the tool name explicitly and get 155 citation works upon \textit{IC3} and 376 for \textit{Epicc}. Then we filter the non-English papers, repeated ones, and degree thesis. Papers that just introduce tools as related works are also removed. 
Totally, we get 48 works that utilize or extend \textit{IC3} and 12 papers for \textit{Epicc}.
Six works are found implementing ICC-analysis modules by themselves instead of using \textit{IC3}/\textit{Epicc} for efficiency or effectiveness reasons. And five works~\cite{DBLP:conf/ccs/BosuLYW17,chen2019storydistiller,DBLP:conf/icse/SamhiBBK21,icse22_iccbot, jhaiccmatt} develop standalone ICC analysis tools.
Moreover, we observe that both the analysis framework \textit{Gator}~\cite{DBLP:conf/kbse/YangZWWYR15} and an early tool\textit{ A$^3$E}~\cite{DBLP:conf/oopsla/AzimN13} provide ATG analysis functionality.

According to the above investigation, ten state-of-the-art tools are discovered, in which ICCMATT~\cite{jhaiccmatt} and RAICC \cite{DBLP:conf/icse/SamhiBBK21} are omitted for requiring source code, and only generating refactored application but not ICC links.
The rest ones are listed as follows. In 2013, \textbf{A$^3$E (2013)}~\cite{DBLP:conf/oopsla/AzimN13} constructs static ATG and uses it to guide the dynamic test generation.
In the same year, \textbf{Epicc (2013)}~\cite{DBLP:conf/uss/OcteauMJBBKT13} reduces the discovery of ICC to an instance of the Inter-procedural Distributive Environment (IDE) problem.
\textbf{IC3 (2015)}~\cite{Octeau15ICSE} is an enhanced tool based on \textit{Epicc}, which uses a generic solver to infer possible values of complex objects in an inter-procedural, flow, and context-sensitive manner. 
\textbf{GATOR (2015)} \cite{DBLP:conf/kbse/YangZWWYR15} is a program analysis toolkit that performs static control-flow analysis on Android apps~\cite{DBLP:conf/icse/YangYWWR15}.
The \textit{ATGClient} is one of its default client provided.
\textbf{IC3DIALDroid (2017)} \cite{DBLP:conf/ccs/BosuLYW17} (IC3$_{Dial}$ for short) extends \textit{IC3} by implementing incremental callback analysis to replace the original one. \textbf{StoryDroid (2019)} \cite{DBLP:conf/icse/ChenFCSLLX19} aims at generating storyboard for apps, which combines the results provided by \textit{IC3} and ICCs extracted with fragments and inner classes features. Another work \textbf{StoryDistiller (2022)}~\cite{chen2019storydistiller} (\textit{StoryD} for short) is an extension of it, which optimized the original tool on both the ATG construction and UI page rendering. 
\textbf{ICCBot (2022)}~\cite{icse22_iccbot} is a code slice and summary-based resolution tool, which considers the modeling of fragment and performs inter-procedural context-sensitive analysis. In the evaluation part, for tools \textit{Epicc} and \textit{StoryDroid}, we only adopt their extended version \textit{IC3} and \textit{StoryDistiller}.

Table~\ref{tab:tools} first gives an overview of the \textit{update time}, \textit{input/output} format, \textit{functionality} of the collected tools.
As some tools are developed by extending others, we list their \textit{base tool} and the fundamental analysis \textit{framework}.
The column \textit{approach} presents the approaches adopted by each tool, in which \textit{Gator} uses a simplified IDE analysis of \textit{Epicc} (according to ~\cite{DBLP:conf/icse/YangYWWR15}), and \textit{StoryD} uses IDE because it first runs \textit{IC3} to get parts of ICCs. 
Then we summarize the analysis sensitivity, including \textbf{f}low, \textbf{c}ontext, f\textbf{i}eld, and \textbf{o}bject sensitivity~\cite{DBLP:journals/infsof/LiBPRBOKT17}, of each tool by investigating their related literature.
Tools \textit{IC3} and \textit{IC3$_{Dial}$} both declared that they use context-sensitive inter-procedural analysis, however, we find several context-insensitive counterexamples in the subsequent evaluations. 
For \textit{StoryD}, we use the same sensitivity with \textit{IC3} because the sensitivity of its own fragment analysis is unknown.
The next column gives the analyzed component type, in which \textit{A$^3$E} and \textit{StoryD} declared that they only construct ATG, but according to our evaluation results, other kinds of components (\textit{NAct}) like services and broadcast receivers are also reported. Overall, only \textit{StoryD} and \textit{ICCBot} concentrate on the analysis of Fragment (\textit{Fr}) component.
The last two columns give whether there are analyses of extra data and string operation.


\subsection{Metrics Adopted by Existing Tools}
According to the evaluation approach presented in the related literature, we summarize the number of hand-made (\#hm) and real-world (\#rw) Android packages (apks) and the evaluation metrics used by each tool.
Note that, \textit{StoryD} has two numbers of \textit{\#hm} and \textit{\#rw} for it has two versions, and \textit{A$^3$E} is not listed as its transition extraction phase is not directly evaluated. 
As shown in Table~\ref{tab:metrics}, the number of identified ICC links (\textit{ICC}) and the ratio of the apps that can successfully pass the analysis without timeout or crash (\textit{succ}) are two popular metrics.
Totally, five tools are evaluated with hand-made apps (labeled with $\bullet$), most of which evaluate the FN and FP ICCs with the labeled ground truths.
Six tools are evaluated with real-world apps (labeled with $\star$). 
However, as no ground truth is available, researchers usually use the number of identified ICC fields that can be determined without uncertainty (\textit{if}), the ratio of covered activity (\textit{cov}), and detected leaks in higher-level analysis (\textit{leak}) as metrics to evaluate the extracted ICCs.

\begin{table}[htbp!]
	\centering
	\setlength{\abovecaptionskip}{0pt}
	\setlength{\belowcaptionskip}{-10pt}
	\footnotesize
	\caption{Evaluation Metrics Adopted By Tools}  \label{tab:metrics}
	\begin{tabular}{|c|c|c|c|c|c|c|c|c|}
		\hline   
		\textbf{Tool} & \textbf{\#hm} & \textbf{\#rw}& \textbf{ICC} &\textbf{succ}  &\textbf{FP/FN} &\textbf{if} &\textbf{cov} &\textbf{leak}\\\hline
		Epicc	& 0 & 1,200 	&		   &    	&   &$\star$   &  & \\ \hline
		IC3		& 0 & 500 		&$\star$   &    	&   &$\star$   &  & \\ \hline
   IC3$_{Dial}$ & 190 & 1,000 	&$\bullet$ $\star$   &$\bullet$ $\star$ &   &   &  & \\ \hline
   		Gator	& 20 & 0		&$\bullet$   &    	&$\bullet$   &   & & \\ \hline
		StoryD  & 10/0 & 50/150 &$\bullet$ $\star$   &    	&$\bullet$   &   &$\star$ & \\ \hline
		RAICC	& 20 & 1,000 	&$\star$   &    	&$\bullet$   &   & &$\star$ \\ \hline
		ICCBot	& 132 & 2,000 	&$\bullet$$\star$   &$\bullet$$\star$ &$\bullet$   &   & & \\ \hline
	\end{tabular} 
	\vspace*{-0.5cm}
\end{table}


%% file: content/dataCollection.tex
\section{Experimental Setup: Data \& Metrics}
This section presents the experimental setup, including the unified framework, data collection process, and metric picking principles.

\begin{figure} [tbp!]
	\setlength{\abovecaptionskip}{0pt}
	\setlength{\belowcaptionskip}{-10pt}
	\includegraphics[width=0.48\textwidth]{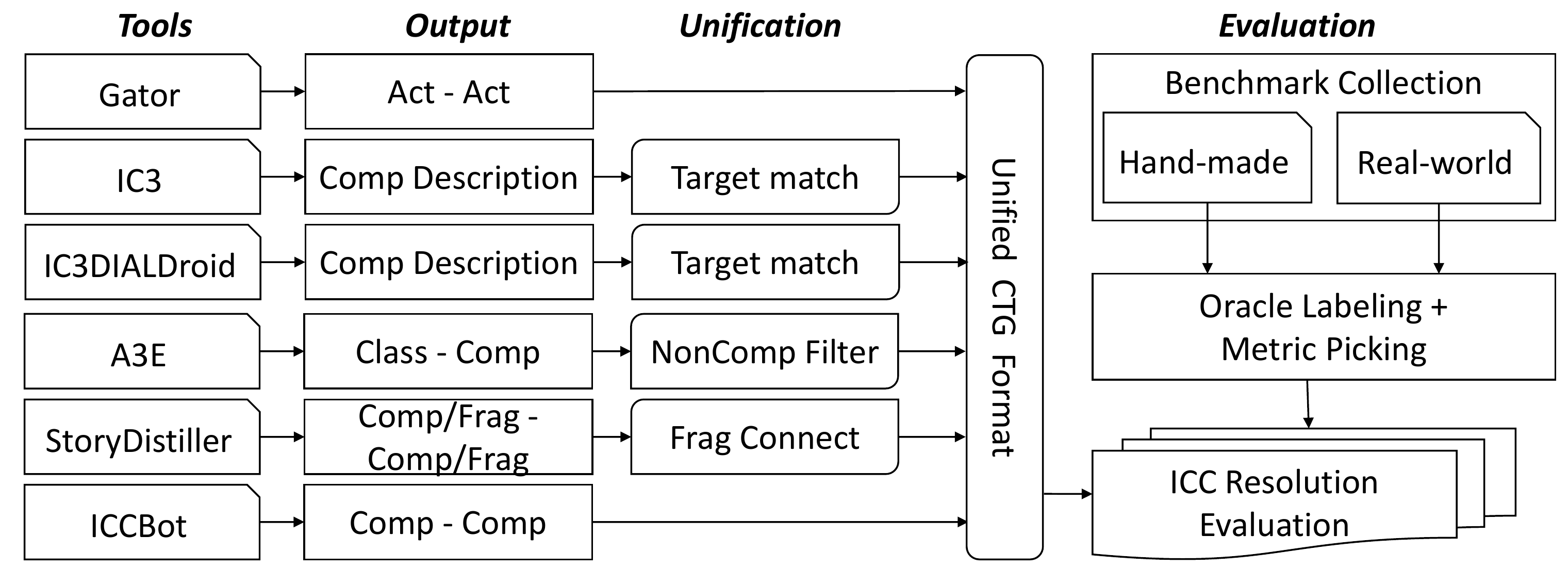}
	\caption{The Unified Evaluation Framework}
	\label{fig:frameWork} 
  \end{figure}

\subsection{Unified Evaluation Framework}
Before evaluation, we noticed that the functionalities and output formats of the state-of-the-art tools vary.
For example, tools \textit{IC3} and \textit{IC3$_{Dial}$} only give the data value of ICC fields instead of the Intent matching results; \textit{A$^3$E} does not filter the ICCs connected with a non-component class; and \textit{StoryD} does not directly connect the components linked by fragments (Act $\rightarrow$ Frg $\rightarrow$ Act). 
To make the comparison possible, we unify the ICC resolution results with several steps: component filtering and enhancing, target matching, and output unifying.
The pre-process of each tool is shown in Fig.~\ref{fig:frameWork}.

\subsection{Multiple-type Benchmark Collection}
In this part, we introduce the collection of multiple-type benchmark suites, including hand-made and real-world apps.

\subsubsection{Test Suites for Hand-made Apps}
For hand-made benchmarks, we prefer the ones used in existing evaluation approaches, which are carefully designed with specific characteristics.
By reviewing the literature that have benchmarks proposed, five hand-made benchmarks with 73 apps are collected as \textbf{BenchHand}.
DroidBench \textbf{(BenD)} \cite{DroidBench} in paper~\cite{Arzt14PLDI} is a benchmark suite that is designed for evaluating the information-flow analyzers. It contains 18 ICC-related apps in its ICC category folder. 
The benchmark ICC-Bench \textbf{(BenI)}~\cite{ICC-Bench} in literature~\cite{DBLP:conf/ccs/WeiROR14} is an ICC-specific benchmark, which contains 24 apps for testing ICC resolution capabilities.
The StoryD-Bench \textbf{(BenS)}~\cite{StoryDroid-Bench} in literature~\cite{DBLP:conf/icse/ChenFCSLLX19} concentrates on two ICC-related characteristics, fragment and inner class, that are not well-addressed by tool \textit{IC3}. It designs 10 test apps around them specifically.
As previous benchmarks do not take the atypical ICC into account, work \cite{DBLP:conf/icse/SamhiBBK21} designed 20 test apps as RAICC-Bench \textbf{(BenR)}~\cite{RAICC-Bench} to supplement \textit{BenD}, in which various atypical ICC usages are provided.
And ICCBotBench \textbf{(BenT)}~\cite{ICCBot} in paper~\cite{icse22_iccbot} is a compact benchmark with one app, which considers various typical usage of fragment loading and data passing among methods.

The ICC-related key characteristics used in these benchmarks can be categorized into several classes,
including the control-flow-related characteristics: \textit{callback entry} and \textit{fragment};
data-flow-related ones: \textit{calling context} and \textit{string operation};
ICC-behavior-related: \textit{implicit match}, \textit{atypical ICC} and \textit{dynamic broadcast receiver};
as well as other-class-related: \textit{library}.
An ICC that does not involve any specific characteristic is labeled as \textit{basic}.
Among them, five characteristics have been introduced in Section~\ref{overview}.
For others, they focus on whether the assignments of ICC resolution related fields are operated by String APIs (\textit{Str Operation}), whether the modeling of specific Android, Java or the third-party library classes is required (\textit{Library}), as well as whether the target broadcast receiver component is declared dynamically (\textit{DynamicBR}).
Table~\ref{tab:benchs} presents the number of related code snippets of each characteristic, in which one code snippet may involve multiple characteristics.

\begin{table}[!tb]
	\centering
	\setlength{\abovecaptionskip}{5pt}
	\setlength{\belowcaptionskip}{-10pt}
	\footnotesize
	\caption{Characteristic-Specific Code in \textit{BenchHand}}\label{tab:benchs}
	\begin{tabular}{|l|c|c|c|c|c|c|}
	\hline   
\textbf{Characteristic}    &\textbf{BenD}	&\textbf{BenI} & \textbf{BenS} 	&\textbf{BenR} 		 &\textbf{BenT} & \textbf{Sum} \\\hline
Basic                       &2             &8		         &15		        &1       &0        &26\\ \hline
Fragment                    &0             &0		         &4	            	&0       &4        &8\\ \hline
Callback Entry		        &0             &16		         &4		            &0	     &11       &31\\ \hline	
Implicit Match	            &5             &13		         &0		            &1	     &9        &28\\ \hline
Calling Context	            &0             	&0		         &15	            &0	     &9        &24\\ \hline
Atypical ICC	            &0             &0		         &0		            &23	     &0        &23\\ \hline	
Library          	        &5              &0		         &0		            &0	     &0        &5\\ \hline
DynamicBR           		&1              &2		         &0		            &0	     &0        &3\\ \hline		
Str Operation  	        	&1              &1		         &0		            &0	     &0        &2\\ \hline
\textbf{Sum}    &\textbf{14}    &\textbf{40}  &\textbf{38}	 &\textbf{25}  &\textbf{33}  &\textbf{150}\\ \hline
	\end{tabular} 
	\vspace{-0.5cm}
\end{table}

\subsubsection{Test Suites for Real-World Apps}\label{benchII}
As existing works perform evaluations on real-world datasets varying from 50 to 2,000 apps, we collect 2,000 open source apps from F-droid~\cite{F-Droid} and 2,000 commercial apps from Google Play~\cite{GooglePlay} as a large-scale benchmark \textbf{BenchLarge}.
All the apps are randomly downloaded from the market \textit{AndroZoo}~\cite{Allix:2016:ACM:2901739.2903508}.
However, it is hard to obtain the complete oracle for up to 4,000 apps. 
Thus, we decide to construct oracles for real-world apps on a compact subset of real-world apps.
Considering the representativeness of test suites, we prefer the apps that suit GUI exploration as well as the ones that may have more ICC links.
First, we pick all the 20 apps that are used in a recent dynamic exploration work~\cite{DBLP:conf/icse/YanLP00L20}, all of which can pass the instrumentation process and are suitable for exploration.
One app is dropped for its source code is unavailable by now, so the ICCs of it are difficult to be confirmed.
Besides, we consider the downloaded apps in \textit{BenchLarge}. For the efficiency of manual auditing with source code, only the F-droid apps are taken into consideration.
We first analyze the number of components in each app and pick the top 40 apps.
Then we filter out the apps that failed the instrumentation and the duplicated ones that are variants of the collected ones.
We also drop the social media apps that require a real identity. 
Finally, we got 31 (19+12) apps (\textbf{BenchSmall}) proper for oracle construction, whose size ranges from 1M to 93M, the average number of GitHub stars is 1,010, and the average number of components is 35.
After collection, we have three benchmark suites with 4,104 apps, including a hand-made one, a large-scale real-world one, and a compact real-world app set.

\subsection{Oracle Construction}
\revise{For the hand-crafted apps in \textbf{BenchHand}, we perform manual review on code to obtain the ground-truth oracle. However,
the specifications for real-world apps in \textbf{BenchSmall} } are not available to the third-party testers, which means the ground truths can not be obtained.
An alternative is to manually collect a subset of real ICCs as an under-approximation of the ground truth for evaluation, which is sound but incomplete.
To guarantee the usability of the constructed oracle, the oracle obtaining approach should be \textbf{practical}, and the \textbf{reliability} of each ICC should be confirmed.

For \textbf{practical} ICC collection, the instrumentation-based dynamic analysis can help to obtain candidate ICCs.
\revise{After inserting method-level probes into apk files, we can automatically collect the runtime information during GUI exploration, and then analyze the component-launching orders from logs.
On one hand, it is applicable for any app that allows apk modification and repackaging.
On the other hand, the dynamically triggered ICCs are not limited by the complexity of static code characteristics, i.e., the corresponding code snippets are with high diversity.
Since dynamic analysis may not be able to trigger all \texttt{intents}, it would introduce bias to the results. 
To improve the overall coverage, we combine the results of state-of-the-art GUI input generation tools and manual exploration. }
First, we utilize the GUI testing tool \textit{APE}~\cite{DBLP:conf/icse/GuSMC0YZLS19} to drive the dynamic execution. 
Each app is explored three times and each execution takes one hour. 
Besides, we manually interact with each app for ten minutes as a supplement.
Overall, the dynamic GUI exploration covers 613/1,103 components, with an average coverage of 58.9\%.

To guarantee the \textbf{reliability} of oracles, we filter the ICC set with two steps.
As the class loading orders could not accurately reflect component transitions, we can not simply use them to build ICCs. Instead, both the lifecycle status of components and the historically visited component stack should be considered.
Besides, many lifecycle methods are not overridden by developers, thus the execution of these methods will not be logged.
Also, components can extend their father classes and invoke their lifecycle methods, which introduces irrelevant components and messes the event order.
Furthermore, there are several types of specific lifecycle behaviors, e.g., launch-mode or flags setting will influence the loading of historically visited components.
For these reasons, we adjust our ICC extraction algorithm to fit these problems, including filtering the polymorphic method invocations, omitting the non-starting callbacks of recently launched components, etc. These works can help us to filter parts of FPs and save labor costs in further code auditing. Details of the Android single- and multiple-component interaction models and the ICC extraction process are displayed along with the collected oracle set in \cite{ICCViewer}.
By the automatic dynamic log analysis, we get 1,339 ICCs as the dynamically constructed oracle.

In the next step, we manually filter the ICCs that cannot be linked to an Intent-sending code snippet and confirm the correctness of 984 ICCs. 
\revise{
First, we globally search both the name of the target component and values of corresponding intent-filters (declared in the manifest file for implicit ICC), by which we can find the ICC sending methods. Then, we trace their callers with the help of the call graph. If we could find a trace starting from a lifecycle method or a callback method, we can finish the search. Note that, for callback methods, we also need to find out how the callback is registered. If the source component is not registered in the manifest file (may be an abstract class), we then review the code of their subclasses. And because many ICCs are passed through fragments, we will search the loaded fragments in the source component and check whether an Intent is sent by a loaded fragment. 
After that, if we still cannot find a code snippet, we filter this ICC out of the oracle set. 
For example, we use the order of callback methods to decide the order of the components, if an ICC is sent by the previously started background services, the source of the ICC may be misidentified.
Besides, the dynamically loaded code may trigger real ICCs during runtime, however, these ICCs cannot be recognized by static ICC resolution tools. 
Other reasons like unmodeled polymorphic relationships and unexpected app restarting also lead to wrongly recognized ICCs during the dynamic analysis. }
Meanwhile, some ICCs that are hidden behind complex control and data flows may be missed.
For example, the longest call trace we successfully tracked involves fourteen method calls and nine classes, including three activities, three fragments, two adapters, etc. It is difficult and time-consuming to confirm ICCs like that.
Besides, during locating the related code snippets for the detected ICCs, we also record the newly observed ICCs for oracle enhancement. 
In this step, 586 ICCs are manually added, after which the number of ICCs is 1,570.

For each extracted ICC, we review its code snippet and point out the typical characteristics involved. As shown in \cite{ICCViewer}, we design 25 code characteristic tags for each ICC and two of the authors label these tags together.
\revise{Table \ref{tab:tags} gives the type and distribution of these 25 tags, involving the type of source or destination components, the entry method of an ICC invocation, how an ICC is sent out, the details of the method calls and Intent field values. }

\begin{table}[tbp!]
	\centering
	\setlength{\abovecaptionskip}{0pt}
	\setlength{\belowcaptionskip}{-5pt}
	\footnotesize
	\caption{Type and Distribution of 25 ICC-related Tags}\label{tab:tags}
	\begin{tabular}{|l|l|}
		\hline   
		\textbf{Type} & \textbf{Distribution}\\\hline
		Component	 	&\tabincell{l}{Activity (96\%), Service (10\%), Broadcast (5\%), Dynamic\\ Broadcast (1\%)}\\\hline
		Non-Component	&Fragment (14\%), Adapter (32\%), Widget (4\%), Other Class (39\%) \\\hline
		Entry Method 	&Lifecycle (79\%), Dynamic (60\%), Implicit (51\%), Static (4\%) \\\hline
		Exit Method		&Normal (94\%), Atypical (6\%) \\\hline
		Method Call 	&\tabincell{l}{Basic (56\%), Callback Listener (53\%), Asynchronous (6\%), \\ Polymorphic (42\%), Library Method (7\%)}\\\hline
		Intent Type		&Explicit Intent (97\%), Implicit Intent (3\%)\\\hline
		Intent Field Value	&\tabincell{l}{Context-related (40\%), Static Value (1\%), Extra Data (40\%),\\ String Operation (0.5\%)}\\\hline
	\end{tabular}
	\vspace{-2em}
\end{table}


\subsection{Metric Picking}
This part discusses the metrics that will be adopted in this study.

\textbf{Oracle Metric.}
For apps with ground truths, the oracle-based metrics true positive (TP), false positive (FP), and false negative (FN) are the best choices, \revise{i.e., which measure whether an ICC identified by the tools has the same source and destination component name with an ICC in the oracle set. }
For apps in \textit{BenchHand}, we can obtain their ground truths by code reviews, and compare their numbers of the TP, FP, and FN ICCs.
For the compact \textit{BenchSmall}, as its under-approximation of the ground truths is extracted, we can get the lower bound of FN ICCs when compared with the labeled oracle.

\textbf{Number Metric.}
In existing works, number-based metrics, e.g., the number of reported ICCs and identified ICC fields, are usually used to evaluate the tools' performance on real-world apps.
The reason is that number-based metrics can reflect the upper bound of the TP ICCs, it is useful when there are few FP ICCs.
However, according to the oracle-based results on \textit{BenchHand}, FPs exist for most ICC resolution tools (refer to Table~\ref{tab:handmade}).
Considering the well versatility on various datasets, we still use number-based metrics to evaluate tools' performance on all three benchmark suites.
Besides, to measure the contribution of the number-based metrics to the ICC resolution results, we take the structure of CTG into consideration, i.e., use the graph-based metrics as a supplement.

\textbf{Graph Metric.}
As the results of ICC resolution can be represented as a directed graph, i.e., the nodes are components and the edges are ICCs, we use the average degree metric to obtain the density of edges.
$deg(CTG) =  2 \times  |E| \div |N|$, in which $|E|$ is the number of reported ICCs and $|N|$ is the number of declared components.
This metric takes both the number of ICCs and the scale of apps into consideration. 
Larger $deg(CTG)$ means more ICCs reported, which can be used to make comparisons among multiple benchmarks.
Besides, we consider the connectivity of the graph with the following three metrics.
The metric $C_{separated}$ denotes the number of isolated components that do not connect to any other;
$C_{mainNot}$ denotes the number of components that are not reachable from the default entry, usually the \texttt{MainActivity};
and $C_{exportNot}$ denotes the number of components that are not reachable from any exported entry component.
From the users' perspective, the lower these metrics, the better CTG connectivity, and more functionalities could be explored.
\revise{There are two possible cases that a component may not be linked to any other component.
One case is, it is an exported component only for external launch. In our dataset, most components (85\%) are either MainActivity or not exported, which should connect to others. 
Meanwhile, many exported activities are not designed for external launch only, i.e., though they can be launched externally, they can also be launched internally, e.g., payment or login activities. 
The other case is dead-code components that are registered but not used, which also rarely happens.
Therefore, we suppose that developers are less likely to design separated or unreachable components in their apps on purpose, and the graph-based metric can work on most scenarios. }

%% file: content/evaluation.tex
\section{Results and Analyses}
\revise{This section aims to answer the following research questions.

\begin{itemize}[leftmargin=0pt]
	\item RQ1: Can existing tools analyze multiple-type apps with high success rate and efficiency?
	\item RQ2: How is the performance of the tools in terms of number \& graph metrics?
	\item RQ3: To what extent can the tools identify ICCs in our oracle set?
\end{itemize}}

\subsection{RQ1: Usability and Efficiency}

First, we explore the configurability of tools. 
For the most popular tool \textit{IC3}, there are seven COAL~\cite{Octeau15ICSE} models that can be configured in its source code. 
However, it brings the extra cost for users to learn the principle and grammar of COAL.
Though both \textit{IC3$_{Dial}$} and \textit{StoryD} extend \textit{IC3}, i.e., are IC3-based tools, they do not modify the inner models.
As \textit{\textit{IC3$_{Dial}$}} optimizes the callback related code snippets and \textit{StoryD} directly invokes \textit{IC3}, they have no extra configuration item.
For \textit{StoryD}, we remove the code related to the dynamic app exploration process and only record the statically extracted ICCs.
Tools \textit{Gator} and \textit{ICCBot} both provide various configuration items.
By inspecting the argument parsing process of \textit{Gator}, we find the ``implicitIntent'' item relates to ICC resolution and set it as true.
For \textit{ICCBot}, we use all its default configurations. 

\begin{figure} [tb!]
	\setlength{\abovecaptionskip}{0pt}
	\setlength{\belowcaptionskip}{-10pt}
	\includegraphics[width=0.47\textwidth]{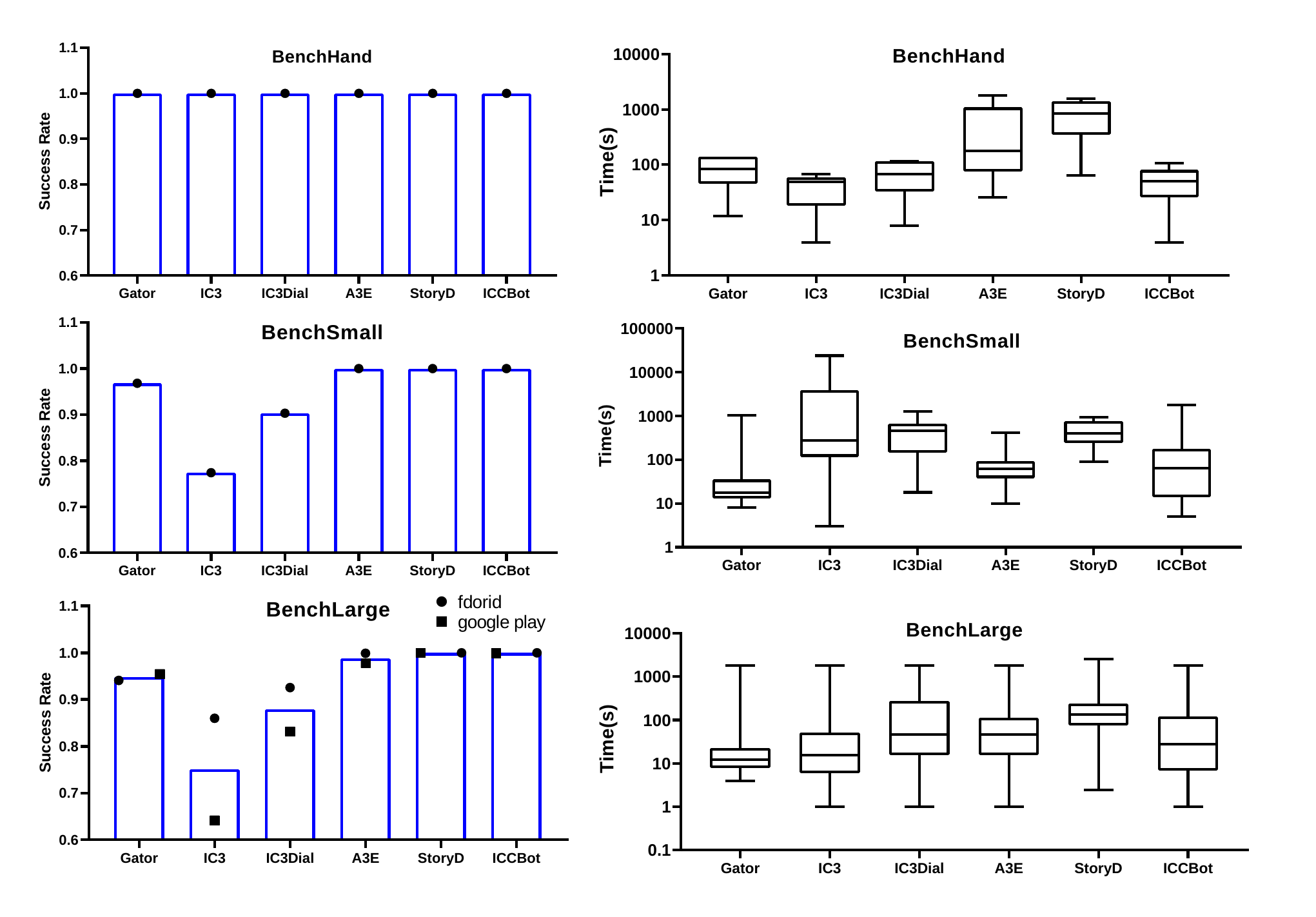}
	\caption{The Success Rate and Execution Time of Tools}
	\label{fig:succe+time} 
  \end{figure}

Then, we compare the success rate during analysis and the execution time of tools.
The whole analysis process is performed on a Linux server with two Intel® Xeon® E5-2680 v4 CPUs and 256 GB of memory.
As shown in Fig.~\ref{fig:succe+time}, on \textit{BenchHand}, all apps can be successfully analyzed within an acceptable time.
On the real-world app set \textit{BenchSmall}, \textit{Gator} outperforms others in efficiency while \textit{IC3-based} tools are more time-consuming, e.g., \textit{IC3} takes more than six hours to analyze app \textit{SuntimesWidget}.
As the users usually invoke fundamental tools in limited time, e.g., 30 minutes in \cite{DBLP:conf/msr/AhmadKSA16}, we use the same setting when analyzing dataset \textit{BenchLarge}.
With such a limit, \textit{IC3} and \textit{IC3$_{Dial}$} suffer from crash or timeout problems and have a lower success rate than others, e.g., they cannot finish analysis for up to 36\% and 17\% google play apps. 
For \textit{StoryD}, though it invokes \textit{IC3}, \revise{it benefits from the 10-minute timeout setting on invoking IC3 and light-weight self enhancement}. 
The different success rates of F-droid and Google Play apps on \textit{BenchLarge} also indicate that the complexity of code can greatly influence the results, and \textbf{the analyzing efficiency on complex real-world apps requires more attention}.
In total, the analysis time for all tools on \textit{BenchHand}, \textit{BenchSmall} and \textit{BenchLarge} are 2, 13 and 892 hours, respectively.

\subsection{RQ2: Use Number \& Graph Metrics}
\begin{figure} [!bp]
	\setlength{\abovecaptionskip}{2pt}
	\setlength{\belowcaptionskip}{-10pt}
	\includegraphics[width=0.48\textwidth]{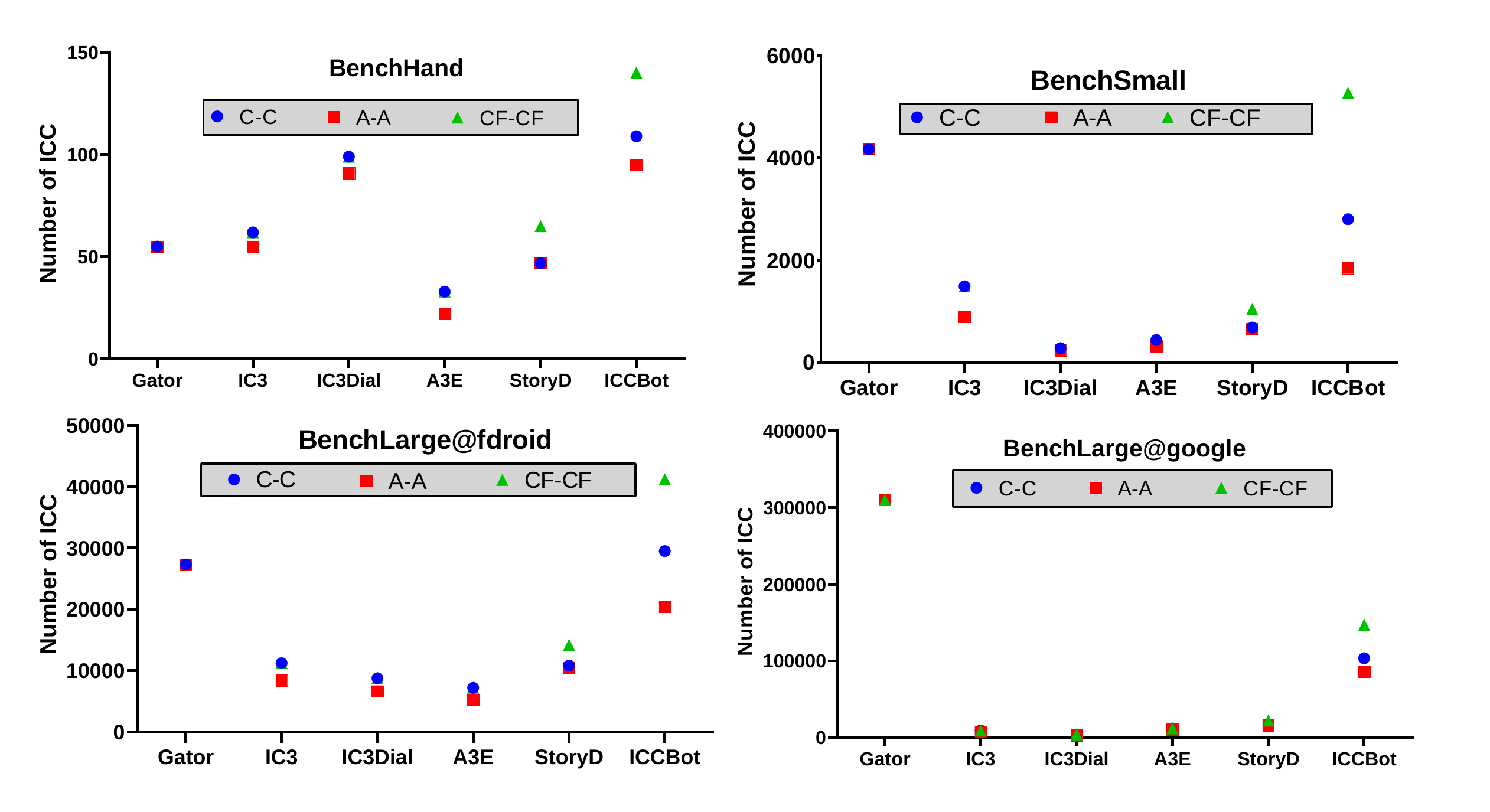}
	\caption{Evaluating with Number-based Metrics}
	\label{fig:EdgeMetric} 
\end{figure}

In Fig.~\ref{fig:EdgeMetric}, we count the number of edges that involve the basic component only (C-C), the activity component only (A-A), and both the basic component and fragment (CF-CF) on each benchmark. Dataset \textit{BenchLarge} is separated into two subsets: F-droid and Google Play set. 
As we can see, the behaviors on \textit{BenchHand} are different from the others, e.g., \textit{IC3$_{Dial}$} generates more ICCs on \textit{BenchHand} while generating fewer ICCs on the other datasets, and the result of \textit{Gator} is the opposite.
The reason is that hand-made apps usually cover the basic usages of one code feature or the combination of a set of features. But it is difficult to design specific code snippets that can cover the FN/FP-related complex patterns that occur in real-world code. 
Thus, \textbf{\textit{BenchHand} is useful to evaluate tools' effectiveness on specific characteristics, while the results are not representative enough due to the differences in code features between hand-made and real-world apps.}


\begin{figure} [!tbp]
	\setlength{\abovecaptionskip}{2pt}
	\setlength{\belowcaptionskip}{-10pt}
	\includegraphics[width=0.48\textwidth]{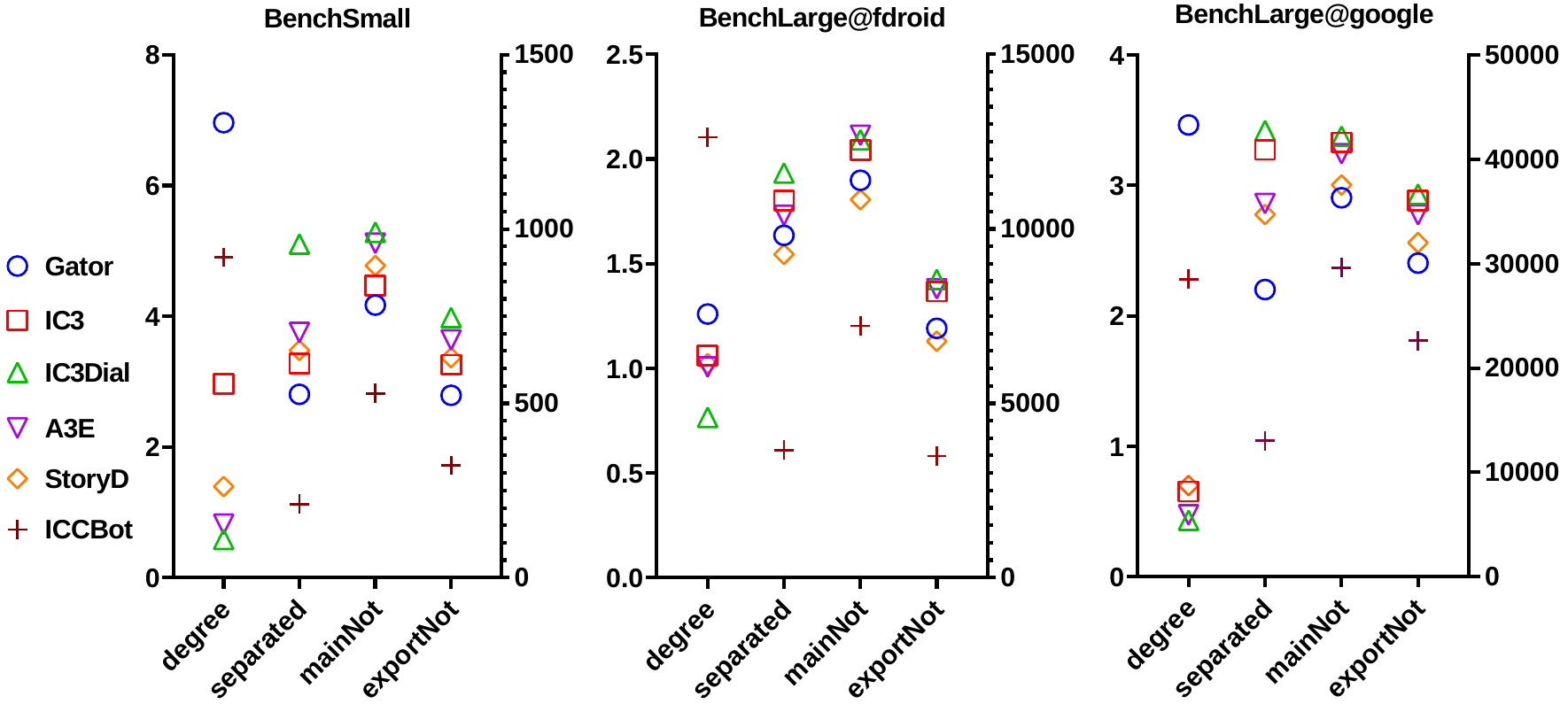}
	\caption{Evaluating with Graph-based Metrics}
	\label{fig:GraphMetric} 
\end{figure}

On all datasets with real-world apps, \textit{Gator} reported the most ICC edges. 
Especially, on the Google Play dataset, it generates more than 300,000 edges, which is 4-80 times more than all others. 
Unfortunately, those results are confusing because we do not know whether they are caused by better analysis ability or higher FP rates.
To figure out that question, we evaluate tools with the graph-based metrics $deg(CTG)$, $C_{separated}$, $C_{mainNot}$ and $C_{exportNot}$ on the constructed CTGs.
The average values of these metrics on three datasets are displayed in Fig.~\ref{fig:GraphMetric}, in which the left Y-axis is for $deg$ $(CTG)$ and the right Y-axis is for others.
Along with a large number of ICCs reported, \textit{Gator} also has high degree values.
However, its connectivity-related values are similar to tools that report much fewer ICCs than it, which means many newly added 
ICCs do not contribute to connectivity improvement. This abnormal behavior guides us to identify many FP ICC candidates in the following Section~\ref{FP}.
Compared to it, tools that have both a relatively high degree and high graph connectivity are more reasonable.
\textbf{In summary, the structure-related information, e.g., the graph-based metrics, can help users notice the unusual behaviors. }


\subsection{RQ3: Use Oracle Metrics}
Both the number and graph metrics can give us an overview of the ICC resolution results.
To obtain more reliable evaluation results, we then measure the effectiveness of tools with oracle metrics.

\begin{table}[!bp]
	\centering
	\setlength{\abovecaptionskip}{5pt}
	\setlength{\belowcaptionskip}{-10pt}
	\footnotesize
	\caption{Evaluating with Oracle on \textit{BenchHand}}\label{tab:handmade}
	\begin{tabular}{|c|c|c|c|c|c|c|c|}
		\hline  
		\multirow{2}{*}{\textbf{Bench}}    &\multirow{2}{*}{\textbf{\#OR}}	&  \multicolumn{6}{c|}{\textbf{\#FP / \#FN}} \\\cline{3-8}
		&&\textbf{\textit{Gator}}	& \textbf{\textit{IC3}}	& \textbf{\textit{IC3$_{Dial}$}} 	
		&\textbf{\textit{A$^3$E}}	&\textbf{\textit{StoryD}}	&\textbf{\textit{ICCBot}} \\\hline
        BenD          	&12		&1/5		&0/4		&0/4		&0/10		&0/9		&0/2 \\ \hline
        BenI	      	&26      &0/22		&0/16		&0/3		&0/19		&0/19		&0/0 \\ \hline
        BenS         	&37      &0/4		&0/4		&0/4		&0/37		&0/1		&0/0 \\ \hline
        BenR		 	&24      &0/24		&0/23		&0/23		&1/1		&0/24		&1/0 \\ \hline
		BenT		   	&11      &3/4		&3/4		&24/1		&0/11		&0/10		&0/0 \\ \hline
        \textbf{Sum}   	&\textbf{110}    &\textbf{4/59}	&\textbf{3/51}	&\textbf{24/35}	&\textbf{1/78}	&\textbf{0/63}	&\textbf{1/2} \\ \hline
	\end{tabular} 
\end{table}

\subsubsection{On BenchHand}
With the labeled oracles, Table~\ref{tab:handmade} gives the evaluation results on benchmark \textit{BenchHand}.
The second column gives the number of ICCs in the oracle set (\textit{OR}) of each benchmark, and the other columns give both the number of FP and FN ICCs.
It shows that FPs happen less often than FNs on most benchmarks, except \textit{BenT}, on which tool \textit{IC3$_{Dial}$} generates many FPs while \textit{IC3} and \textit{Gator} also generate a few ones.
By reviewing these FP-related reports, we find that the reachability analysis of methods and the context value tracking results affect the results.
\revise{For the reachability computing, some tools first compute the methods that could reach an Intent-sending statement, and then compute the data values that could be assigned to that Intent object.
During the two-step reachability computing, the relationship between these context values and the method calls is omitted. 
For example, in the case study in Figure~\ref{fig:FP2}, the decorator method may be reused by multiple callers under various contexts, which leads to many FPs by tools. }
Many ICCs are missed on \textit{BenR} as it contains various atypical types of ICC usage, which requires the modeling of specific APIs.
\textit{BenI} also leads to many FNs because it uses several not commonly used callback methods.
Overall, on \textit{BenchHand}, \textit{StoryD} and \textit{A$^3$E} behave well on FP rate, \textit{ICCBot} behaves well on both FP and FN rates.

Based on the labeled information, we further compare the distribution of code characteristics of FN ICCs.
Compared with other tools, both \textit{ICCBot} and \textit{A$^3$E} work well with characteristic \textit{atypical ICC}, as \textit{ICCBot} adds atypical APIs into the Intent model while \textit{A$^3$E} simply reports ICCs if the creation of an Intent object is identified.
Both \textit{A$^3$E} and \textit{StoryD} fail to identify ICCs with implicit Intent.
According to Fig.~\ref{fig:frameWork}, they both generate CTG directly but do not apply implicit matching, while others consider it by themselves or by the target matching module in our unified framework. 
Thus, though \textit{StoryD} invokes \textit{IC3}, it has more FNs than \textit{IC3} on some cases.
Compared to others, \textit{A$^3$E} is the only tool that failed to resolve all the \textit{calling context} related ICCs.


\subsubsection{On BenchSmall}
\begin{figure} [!bp]
	\setlength{\abovecaptionskip}{0pt}
	\setlength{\belowcaptionskip}{-10pt}
	\includegraphics[width=0.48\textwidth]{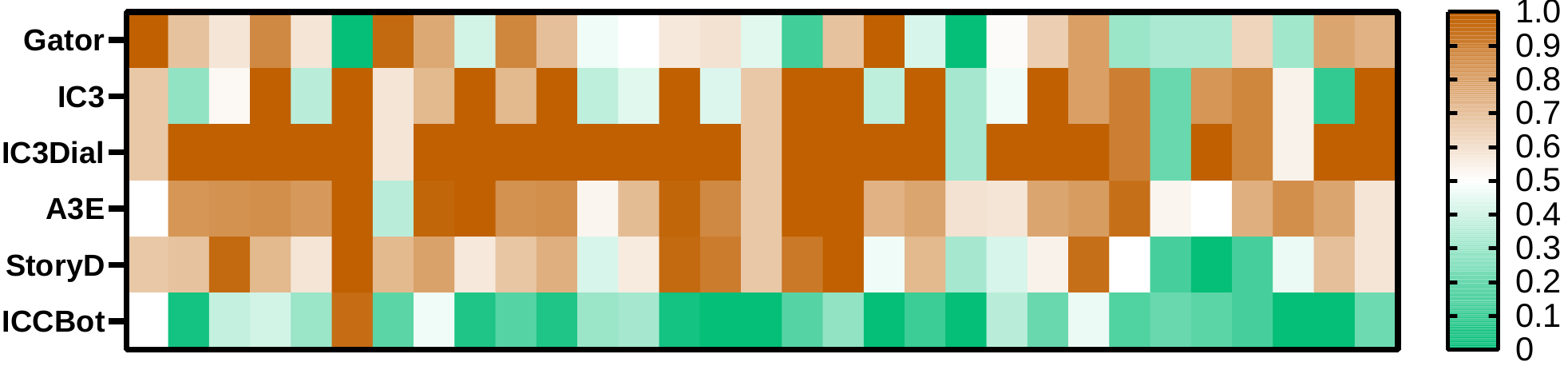}
	\caption{False Negative Rates of Tools on \textit{HandSmall}}
	\label{fig:fnprr} 
\end{figure}

\begin{figure} [bp!]
	\setlength{\abovecaptionskip}{0pt}
	\setlength{\belowcaptionskip}{-10pt}
	\includegraphics[width=0.48\textwidth]{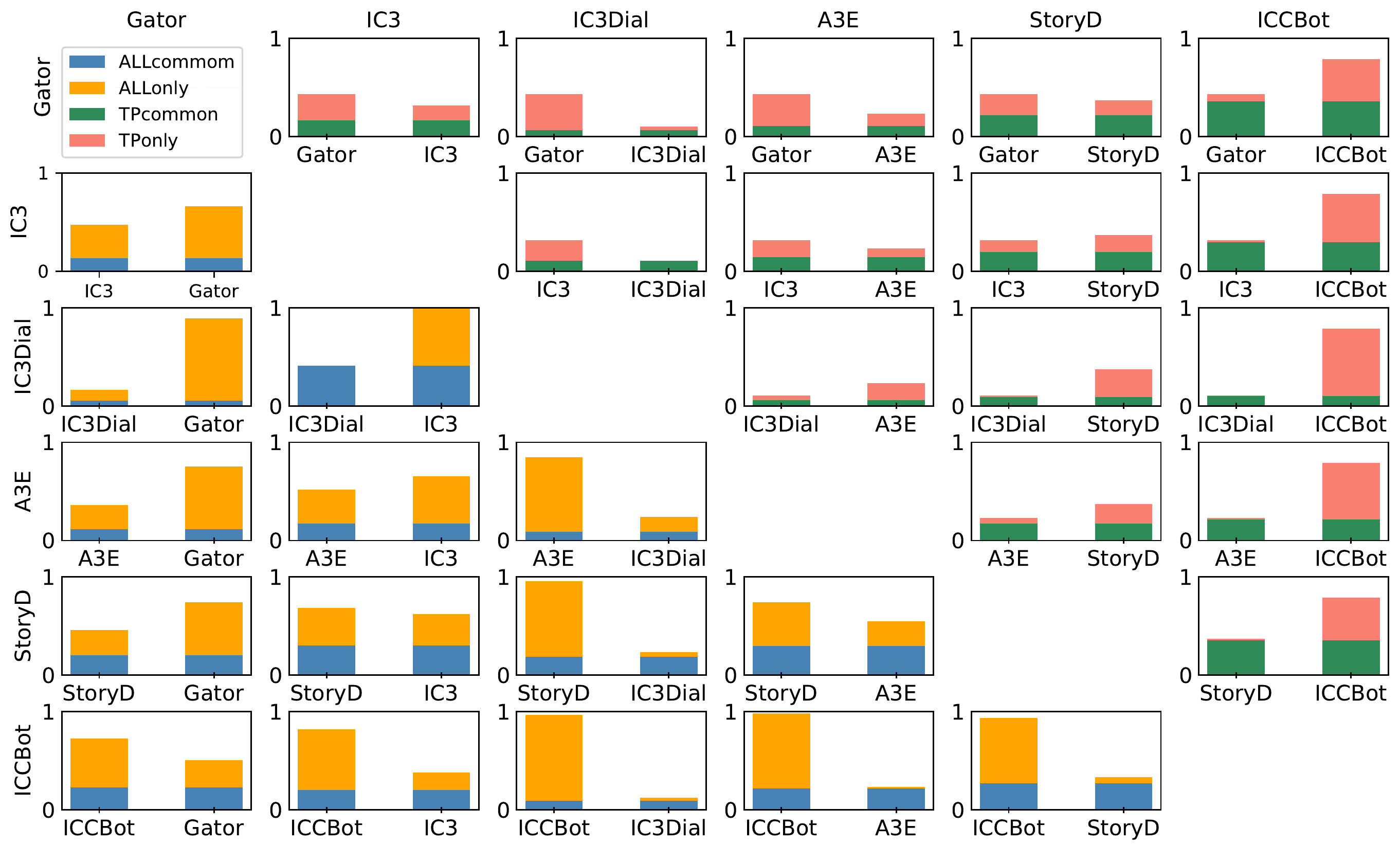}
	\caption{Pairwise Comparison about Reported and TP ICCs}
	\label{fig:pairwiseComparison} 
\end{figure}

Fig~\ref{fig:fnprr} gives the hot-map graph of FN rate results on \textit{BenchSmall}, in which each unit square denotes the FN rate on one app, and the X-axis displays the 31 apps that are sorted by the number of ICCs in the oracle set.
Using reliable oracles with 1,570 edges, we find that around 38\% to 85\% ICCs are missed by the six picked tools, and their average FN rates on apps vary from 21\% to 88\%. 
\textbf{That is to say, there are still massive FN ICCs when working on real-world apps.}
Then, we perform pairwise comparisons to figure out the common and unique ICCs reported by tools. 
In Fig.~\ref{fig:pairwiseComparison}, the bottom left figures are about all the reported ICCs, whose Y-axis values are the ratios of reported ICCs in the union of ICCs reported by two tools.
And the upper right ones only count the TP ICCs, whose Y-axis values are the ratios of TP ICCs in the oracle ICC set.
As we can see, \textit{IC3} covers all the reported and the TP ICCs of \textit{IC3$_{Dial}$} on this benchmark, and the results of other tools are all overlapped.
For instance, even though \textit{ICCBot} can cover most TPs reported by others, every other tool can still report a few TP ICCs that are missed by it.
Furthermore, the union of any two tools cannot cover all ICCs in the oracle set.

\begin{figure} [thbp!]
	\setlength{\abovecaptionskip}{0pt}
	\setlength{\belowcaptionskip}{-10pt}
	\includegraphics[width=0.48\textwidth]{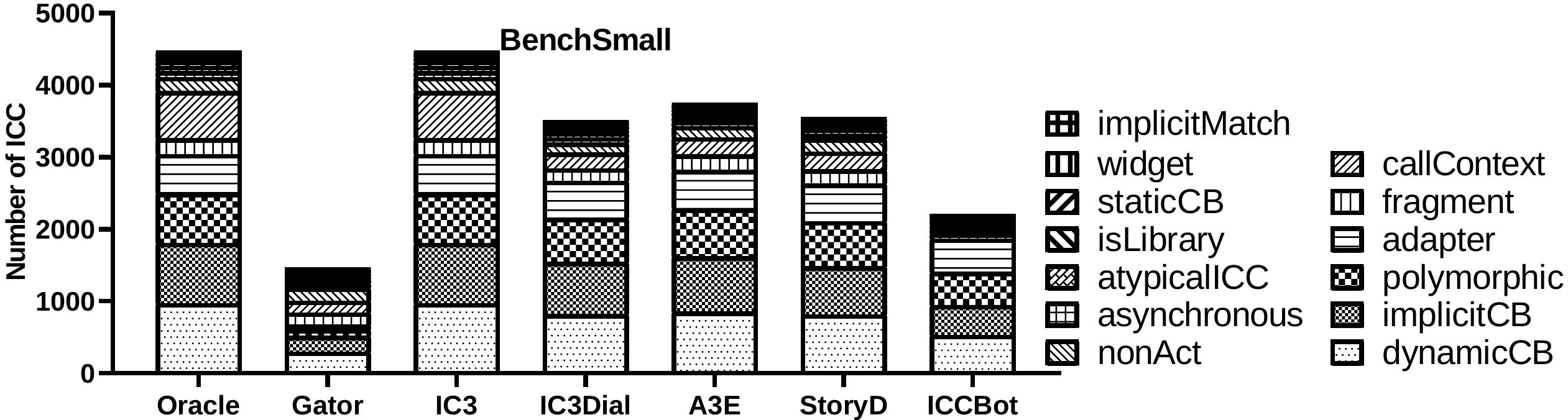}
	\caption{Characteristics of FN ICCs on \textit{BenchSmall}}
	\label{fig:tagprr2} 
	\vspace{-1em}
\end{figure}

Fig.~\ref{fig:tagprr2} presents the top FN-related characteristics of each tool on \textit{BenchSmall}, in which the left bar shows the characteristics of all the ICCs in the oracle set, and others are about the FN ICCs of each tool.
Note again that one ICC may have multiple characteristic labels and the failed apps are not counted for each tool.
According to the results, the \textit{callback entry} related, especially the dynamic and implicit callbacks (CB), ICCs and FNs are both on a large scale.
One reason is, callback entry identification is a big challenge due to the various forms of entry declaration.
Moreover, many other characteristics show up together with callback characteristics, so they may be repeatedly counted.
Compared with the results on hand-made apps, the non-basic-component related characteristics are popular, including the use of \textit{fragment}, \textit{adapter}, etc., which means that \textbf{the ICC sending procedure in real-world code is much more complex than in hand-made snippets}.
The Java-specific characteristics \textit{polymorphic} and \textit{asynchronous} have great influence on the method control flow. Many ICCs related to them failed to be extracted.
And characteristics like \textit{string operation} and \textit{dynamic broadcast receiver} are not counted because few FNs relate to them, in which string manipulations are more often used in malicious apps but only benign apps are picked in our study.

\begin{figure*} [htbp!]
	\setlength{\abovecaptionskip}{0pt}
	\setlength{\belowcaptionskip}{-10pt}
	\includegraphics[width=0.95\textwidth]{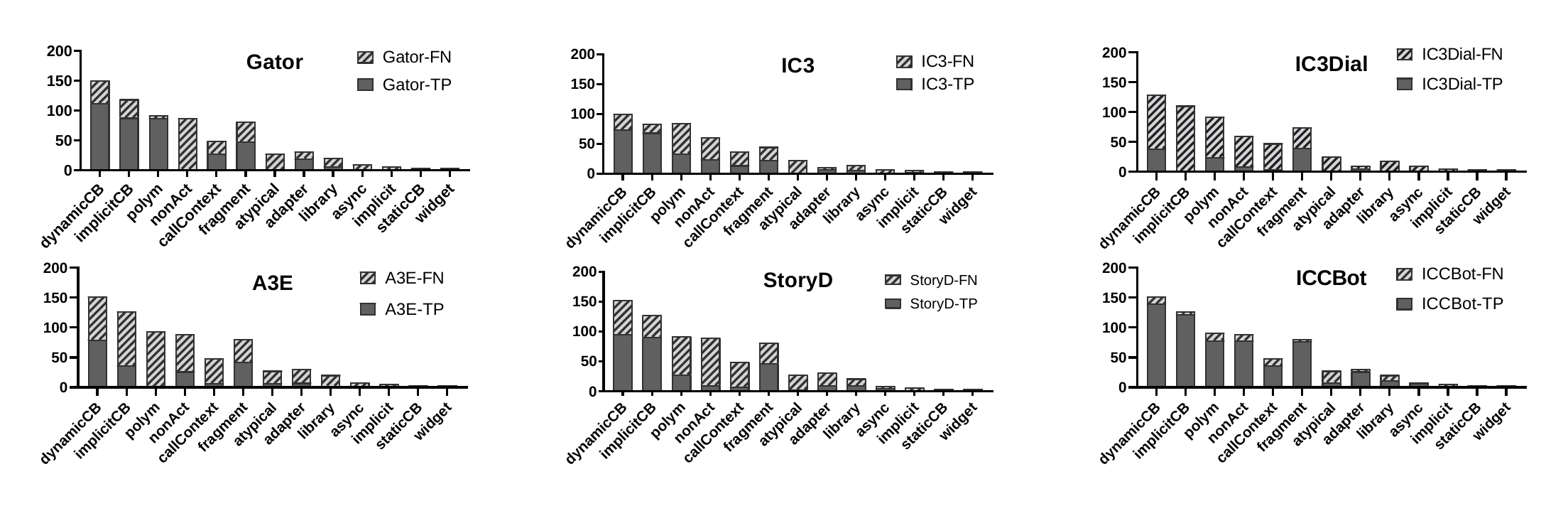}
	\caption{Number of FN and TP ICCs related to Single Characteristic on \textit{BenchSmall}}
	\label{fig:controlvarprr} 
\end{figure*}

To avoid the mutual influences among characteristics, we take each characteristic as a separated control variable and count the ICCs that are only related to it.
For callback-related ones, we pick ICCs that only satisfy one callback type but are not labeled with any other characteristics.
For other characteristics, we omit their callback setting because most ICCs relate to callbacks.
As the atypical ICCs are usually related to non-activity components, their component types are not limited.
The FN and TP results of each tool are shown in Fig.~\ref{fig:controlvarprr}.
As we can see, the Java-specific characteristic \textit{asynchronous} is only concerned by two tools, \textit{ICCBot} and \textit{StoryD}, and characteristic \textit{polymorphic} is omitted by tool \textit{A$^3$E}. Though \textit{static callback} is not a new Android feature, \textit{IC3}, \textit{IC3$_{Dial}$} and \textit{A$^3$E} all fail with the characteristic.
Among all characteristics, three have tight relationships with the evolution of the Android framework, including \textit{implicit callback}, \textit{fragment}, and \textit{atypical ICC}, which are called newly-introduced characteristics, and others are pre-existing ones. 
By evaluating the number of ICCs influenced by each single characteristic, we find that more missed ICCs are caused by the inadequate analysis of pre-existing characteristics (73\%) than the newly-introduced ones (27\%). So, even without consideration of the evolution of the Android framework, \textbf{there is still a long way to go to improve the precision of ICC resolution tools}.



\subsection{Observations for Further Improvement}
According to the above evaluation results, we have the following observations worthy to be discussed.

\textbf{\textit{(For Tool Developer)} }
\textbf{First, the standard evaluation benchmark suites and suitable metrics for fundamental analysis modules are required.} 
As we can see, there are great differences between self-made and complex commercial codes. For further tools that work on ICC resolution, developers could reuse the datasets and metrics provided in this paper. For other problems, developers can leverage the dynamic information to help the benchmark construction for static tools, and vice versa.
Moreover, as the oracles on the complex dataset are not available, it may be helpful to utilize the structure-properties of results, e.g., consider the design intention of CTGs when evaluating ICC transitions, in evaluation.

\textbf{\textit{(For Tool Developer)} }
\textbf{Second, the efficiency of static analyzers should raise more attention.} 
Efficiency and efficiency-induced execution failures usually trouble users \cite{DBLP:conf/icse/TsutanoBSRD17,DBLP:journals/tdsc/ChangLSWJHW21}.
According to our evaluation, the efficiency on complex real-world apps is not satisfactory for most tools, as some expensive analysis approaches may bring unpredictable time costs for users and may not bring equivalent benefits. 
A simple but practical strategy is to store the intermediate results during analysis and allow users to see the partial results at a certain time point. Besides, how to effectively distribute computation resources during analysis should be further explored. For example, developers can make a quick scan of code to decide the order of analysis units, e.g., class or method. Or they can dynamically evaluate the time cost of each analysis unit and handle the costly ones specifically.

\textbf{\textit{(For Tool User)} }
\textbf{Third, concerning more about the trustworthy analysis chain.} 
Many high-level analyses rely on the ICC resolution results, and ICC resolution also relies on the precision of other modules. As the imprecision in the low-level analysis may be propagated to the higher level, the imprecision in low-level tools may greatly influence the final performance with unclear root causes.
Thus, users should get a comprehensive look at any invoked tools to build a trustworthy analysis chain.
And more experimental researchs should be performed to give a many-sided overview of various fundamental analysis tools.

\textbf{\textit{(For Tool User)} }
\textbf{Forth, keeping aware of your key requirements.} 
According to the results, numerous ICCs are missed by six state-of-the-art tools, which means there is not a perfect solution to resolve ICCs from complex real-world apps. 
Therefore, based on our comprehensive evaluation results, users should make decisions according to their key requirements.
Here, we give a group of possible requirements and the corresponding candidate tools in Table~\ref{tab:requirements}, involving the efficiency, completeness, soundness, scenes to be used, and key characteristics concerned. 
For example, \textit{ICCBot} works well in terms of both efficiency and effectiveness and supports all types of components, which can be directly used for CTG construction.
\textit{Gator} outperforms other tools in analysis efficiency so that it can be used in time-conscious scenarios.
For the updating of IC3-based tools, \textit{StoryD} and \textit{IC3$_{Dial}$} can be adopted as they concern fragment and entry-point identification, respectively.
Note that \textit{IC3$_{Dial}$} only works well on parts of apps and the reason is unclear.
\textit{StoryD} provides dynamic UI reference and \textit{Gator} has a static UI analysis client, so they can combine with layout analysis.
And if users pay attention to the data carried with ICC, they can try \textit{ICCBot}, \textit{IC3}, \textit{Gator}, which have such Intent-field analysis.


\begin{table}[htbp!]
	\centering
	\setlength{\abovecaptionskip}{5pt}
	\setlength{\belowcaptionskip}{-10pt}
	\footnotesize
	\caption{Candidate Tools with Key Requirements} \label{tab:requirements}
	\begin{tabular}{|l|l|}
		\hline   
		\textbf{Requirement \& Candidate Tools} & \textbf{Requirement \& Candidate Tools}\\\hline
		Less time cost $\rightarrow$ \textit{Gator, A$^3$E, ICCBot} 		&IC3-based update  $\rightarrow$  \textit{StoryD, IC3Dail, IC3} \\\hline
		More real ICCs  $\rightarrow$ \textit{ICCBot, Gator, StoryD} 		&UI analysis $\rightarrow$ \textit{StoryD, Gator} 	\\\hline
		Fewer fake ICCs  $\rightarrow$ \textit{ICCBot, StoryD, IC3} 			&Intent field extract $\rightarrow$ \textit{ICCBot, IC3, Gator} 	\\\hline
		High SuccRate  $\rightarrow$ \textit{ICCBot, StoryD, A$^3$E}		&Fragment-aware $\rightarrow$ \textit{ICCBot, StoryD} \\\hline
	\end{tabular} 
	\vspace{-2em}
\end{table}

%% file: content/patterns.tex
\section{Root Causes and Patterns}


\subsection{False Negatives in ICC Resolution}
\revise{As shown in the pairwise comparison results between tools, tools have their specific FN ICC sets. By comparing their differences, we separate FN ICCs into two categories: missed by parts of tools and missed by all tools. 
For the ICCs missed by parts of tools, one reason is the lack of analysis on specific characteristics, e.g., the omission of \textit{fragment} by Gator.
Meanwhile, the efficiency of the analysis approach is another reason.
Many ICCs are missed because some tools cannot finish the analysis within a given time, e.g., \textit{IC3} reaches timeout on many apps.
By comparing the FN set of tools, we find that 158 ICCs are missed by all tools. 
Then we analyze the value of the 25 labeled tags of these 158 ICCs, compare the distribution of tags on these ICCs and on all ICCs (refer to Table 4), and find several tags have a higher ratio on the commonly missed ICCs, including fragment, static callback, etc. 
Based on the ICC triggering path labeled in our dataset, two of the authors discuss how can a specific tag characteristic influence the identification of ICC. Finally, we find 26 ICCs are layout-related, 75 involve multiple callbacks, 49 for inter-procedural assignment, 26 are about container-modeling, and we also find 5 special cases caused by implicit assignment and record it. 
Five common FN patterns are discussed as follows.}

\textbf{\textit{P1: Layout-related Callback.}}
There are several forms of callback entries related to XML layout files.
The first line in Listing~\ref{fig:FN1} gives the normal type of static callback, which declares a callback for a \texttt{button} widget. 
Following, a customized view \texttt{navView} is statically declared, which indeed has a dynamic callback in \texttt{navView.class} and will send ICC in this callback method.
Besides, the \texttt{PreferenceScreen} provided by the Android framework supports another implicit way to trigger Intents. We list one of its usage here.
All these patterns require the analysis of layout files.

\begin{lstlisting}[caption={FN: Layout-related Callback}, label={fig:FN1}]
//In the layout file of Component A.class (A to Tgt)
<Button android:id="@+id/button" android:onClick="onMyClick" />
<com.pkg.navView android:id="@+id/navView" /> 
//In com.pkg.navView.class
setNavigationItemSelectedListener(new OnNavigationItemSelectedListener(){ 
    public void onNavigationItemSelected(View v){ 
        startActivity(new Intent(com.pkg.Tgt.class)); }});
//In Component B.class and its layout file (B to Tgt)
public void onCreate(Bundle savedInstanceState) {
    addPreferencesFromResource(R.xml.item); }
<Preference android:key="target"> //one Preference in the PreferenceScreen
    <intent android:targetPackage="com.pkg" android:targetClass="com.pkg.Tgt"/>
</Preference>
\end{lstlisting}

\textbf{\textit{P2: Multi-step Callback.}}
In some cases, the callback recognition requires multiple analyzing steps.
This type of design is commonly used, e.g., reach a view that may trigger Intent sending after clicking another widget.
In Listing~\ref{fig:FN2}, the callback \texttt{onDrawerOpened()} is hidden behind the callback \texttt{onClick()} and the asynchronous method \texttt{run()}.
This pattern requires precise call graph construction as well as a multiple-turn callback analysis.

\begin{lstlisting}[caption={FN: Multi-step Callback}, label={fig:FN2}]
// In Component A.class (A to Tgt)
pendingRunnable = new Runnable() {
    public void run() {
        addListener(new DrawerToggle(){ 
            public void onDrawerOpened(View v){ 
                startActivity(new Intent(Tgt.class));}});
button.setOnClickListener(new OnClickListener(){ 
    public void onClick(View v){ 
        new Handler().post(pendingRunnable); }});
\end{lstlisting}

\textbf{\textit{P3: Inter-procedural Assignment.}}
In Listing~\ref{fig:FN3}, the Intent object is obtained from the return value of method \texttt{getIntentObj()}.
For inter-procedural assignments, the passed value can be parameters, return values, and even the static variables. 
Note that, without tracking a global path, it is difficult to get the precise value of static variables.
For others, careful inter-procedural analysis is required.

\begin{lstlisting}[caption={FN: Inter-procedural Assignment}, label={fig:FN3}]
// In Component A.class (A to Tgt)
public void onCreate(){
     startActivity(B.getIntentObj(A.this)); }
// In Component B.class
public static Intent getIntentObj(Context ctx){ 
    return new Intent(ctx, Tgt.class); }
\end{lstlisting}

\textbf{\textit{P4: Container Modeling.}}
\texttt{Adapter} is a widely used data container that is not well-modeled by now.
Not only the constant data can be stored in adapters, but fragment instances can also be added to it.
The combination of adapter and fragment is popular when using \texttt{ViewPager} component, which is used to switch views according to user operation and each view can be a fragment contained in the adapter.
Like Listing~\ref{fig:FN5} shows, component \texttt{A} loads fragment \texttt{F}, whose instance is stored in \texttt{mAdapter}. And the fragment \texttt{F} launches component \texttt{Tgt} when attached. 
Without the modeling of adapter operating APIs, we can not figure out which fragment is loaded here.
Besides multiple types of adapters, there are also various types of data containers, whose modeling is a challenge to both the control flow edge and data value extraction.

\begin{lstlisting}[caption={FN: Container Modeling}, label={fig:FN5}]
// In Component A.class (A to Tgt)
public void onCreate(){
    mViewPager.setAdapter(mAdapter); 
    mAdapter = new FragmentPagerAdapter( getSupportFragmentManager()) {
        public Fragment getItem(int position) {
            switch (position){ case 0:  return new F();  ...}}};}
// In Fragment F.class
public void onAttach(Activity act) { 
    startActivity(new Intent(Tgt.class));}
\end{lstlisting}

\textbf{\textit{P5: Implicit Assignment.}}
In Listing~\ref{fig:FN4}, component \texttt{A} first loads fragment \texttt{F}.
As the fragment \texttt{F} is attached, it invokes the method \texttt{onDoAction()} in component \texttt{A}, which triggers the ICC $A \rightarrow Tgt$.
However, to detect it, we have to know the actual value of the parameter \texttt{act}. 
In the Android framework, the parameter of \texttt{onAttach()} equals the host Activity of the current fragment, which is an implicit data assignment.
Thus, besides the modeling of the fragment loading behaviors, it also requires modeling the implicit data relationships like this.

\begin{lstlisting}[caption={FN: Implicit Assignment}, label={fig:FN4}]
// In Component A.class  (A to Tgt)
public void onCreate() { 
    loadFragment(F.class); }
public void onDoAction() { 
    launchActivity(new Intent(this, Tgt.class);); }
// In Fragment F.class
public void onAttach(Activity act) {
    ((OnDoActionListener) act).onDoAction(); }
\end{lstlisting}

\subsection{False Positives in ICC Resolution} \label{FP}
\revise{To find out the possible FPs, we pick up apps with the highest value of $deg(CTG)$ for investigation ($deg(CTG)$>15), including \textit{OpenKeychain} (Gator, 19.0), \textit{easydiary} (Gator, 25.4), \textit{SuntimesWidget} (IC3, 22.5), etc.
For these apps, we carefully read their code and infer why a nonexistent ICC is reported. The process is the same as how we identified the correctness of ICC during the dynamic analysis (refer to Section 3). 
After that process, there are still some ICCs failed to be confirmed.
For these cases, we try to infer why a nonexistent ICC is reported as tools do not provide details about why they report such an ICC.
For the possible patterns, we also construct test cases to verify whether an inferred FP pattern can indeed lead to FPs or not. 
The final three patterns we observed (P6-P8) are all verified. }
Through experiments, we also find the simplified model will lead to FPs. 
For example, \textit{A$^3$E} only identifies the Intent declaration statements but not the complete behavior of Intents, so that fake Intents that are not really sent out are reported.
\textit{ICCBot} tries to track the entry points of ICC. For the complex callback registrations that are missed, it takes the top method that it could track as the entry method, while sometimes this simplification brings errors.
Finally, we summarize three concrete circumstances that lead to FPs.


\textbf{\textit{P6: Polymorphic Invocation.}}
The invocation relationships become complex when encountering the \textit{polymorphic} characteristic.
In Listing~\ref{fig:FP3}, subclasses \texttt{SonA}, \texttt{SonB} and \texttt{SonC} all extend class \texttt{Father} and implement the abstract method \texttt{fatherMethod()}, which is invoked in method \texttt{onCreate()}.
Obviously, there are two ICCs, i.e., \texttt{SonA} launches \texttt{SonB}, and \texttt{SonB} launches \texttt{Tgt}. 
However, \textit{IC3} reports four ICCs, and \textit{Gator} reports six ones.
They both compute the reachability between the Intent sending statements and basic component classes, in which the reachability depends on the precision of the call graph.
When combined with the \textit{polymorphic} characteristic, the invocation of a method depends on the execution context, the omitting of which will wrongly connect methods and raise incorrect ICCs.
In this example, tools take all the implementations of \texttt{fatherMethod()} in the same way, which leads to fake call edges.
Moreover, this problem is unexpectedly expanded for \textit{Gator}. In \texttt{SonB}, method \texttt{getIntent()} is invoked to receive Intent from outside. Without object-sensitive analysis, \textit{Gator} misidentifies two Intent objects and takes all the possible sources of \texttt{SonB} as the source ICC being sent out, including the FP sources \texttt{SonB} and \texttt{SonC}. 
The transitivity of FPs may cause exponential growth of ICC numbers.

\begin{lstlisting}[caption={FP: Polymorphic Invocation}, label={fig:FP3}]
// In abstract class Father.class
public void onCreate() { super.onCreate(); fatherMethod(); }
abstract public void fatherMethod();
// In Activity SonA.class (SonA extends Father)
public void onCreate() { super.onCreate(); }
public void fatherMethod() { 
    startActivity(new Intent(this, SonB.class)); }
// In Activity SonB.class (SonB extends Father)
public void onCreate() { super.onCreate();
    Intent received = getIntent();
    startActivity(new Intent(this, Tgt.class)); } 
public void fatherMethod(){ /** do nothing **/} 
// In Activity SonC.class (SonC extends Father)
public void onCreate() { super.onCreate(); }
public void fatherMethod(){ /** do nothing **/ }    
\end{lstlisting}

\textbf{\textit{P7: Decorator Method.}}
In Listing~\ref{fig:FP2}, method \texttt{launchAct()} is a decorator method that invokes the API \texttt{startActivity()} and adds Intent flags for it.
Both components \texttt{A} and \texttt{B} invoke the method \texttt{launchAct()} and pass an Intent object to it.
However, both the \textit{IC3}-based tools and \textit{Gator} adopt context-insensitive analysis for decorator methods, which means the possible targets for \texttt{launchAct()} are extracted from all the received Intents and the sources are all the caller components.
In this case, components \texttt{A} and \texttt{B} are the sources, \texttt{C} and \texttt{D} are the targets, i.e., all the four ICCs will be reported, in which two of them ($A \rightarrow D$, $B \rightarrow C$) are FPs.

\begin{lstlisting}[caption={FP: Decorator Method}, label={fig:FP2}] 
// In Class Util.class
public static void launchAct(Context ctx, Intent i) { 
    addFlagForIntent(i); ctx.startActivity(i);}
// In Component A.class
public void onCreate(){ Util.launchAct(new Intent(getBaseContext(), C.class));}
// In Component B.class
public void onCreate(){ Util.launchAct(new Intent(getBaseContext(), D.class));}
\end{lstlisting}

\textbf{\textit{P8: Type Inference.} }
In Listing~\ref{fig:FP1}, component \texttt{A} dynamically rigisters two broadcast receivers and set corresponding intent-filters for them.
Then, it sends a broadcast with the action value ``\texttt{FilterA}'', which should be received by the instance \texttt{br1} of class \texttt{Receiver1}.
However, in \textit{IC3}, both \texttt{Receiver1} and \texttt{Receiver2} are labeled as the receiving target classes.
By debugging, we find that \textit{IC3} failed to track the correct type of \texttt{br1} and \texttt{br2} for they are field variables. By a conservative analysis, it takes all the broadcast receivers in the app as the target types for registration, i.e, the two \texttt{intent-filters} are registered to both receiver types. 
Without carefully concentrating on the scope of variables and the type inference, FP ICCs can be wrongly reported.

\begin{lstlisting}[caption={FP: Type Inference}, label={fig:FP1}]
public class A extends Activity { // In Component A.class 
    BroadcastReceiver br1, br2;
    public void onCreate(Bundle savedInstanceState) {       
        br1 = new Receiver1();   
        br2 = new Receiver2();
        registerReceiver(br1, new IntentFilter("FilterA"));
        registerReceiver(br2, new IntentFilter("FilterB")); 
        sendBroadcast(new Intent("FilterA"));  ...}}  
\end{lstlisting}

\subsection{Handling of FN/FP Patterns}
\revise{Among the five FN patterns, both patterns P1 and P2 are callback-related. Meanwhile, the identification of a single callback also leads to many FNs.
Tool developers could extend their callback identification module to handle these specific cases, for which the key challenge is how to automatically identify the layout-related and user-customized callbacks precisely.
Pattern P3 depends on whether the analysis approach is path- and context-sensitive. The handling of this pattern is related to the design of the tool and may need more effort.
Besides these patterns, the atypical ICC leads to FNs on many tools. Developers could quickly update the exit method set to support this characteristic.
It is also not hard to extend tools to support non-Activity components, e.g., Service. However, performing extension around fragment, container (P4), and inter-procedural assignment (P5) is not easy and requires fine-grained models.
For the FP patterns, developers could adopt more precise call graph and type inference analysis algorithm to avoid P6 and P8, while to reduce the FPs related to P7, context-sensitive analysis is required.
}

%% file: content/discuss.tex
\section{Threats to Validity}
In this section, we discuss the threats to the validity faced by this work.
The threats to external validity relate to the generalizability of the experimental results. 
Our oracles for real-world apps are extracted from 31 benign Android projects on the public markets, while the results may not generalize beyond the 31 apps, especially the malicious apps.
Threats to internal validity concern factors internal to our approach.
We manually confirm the correctness of the dynamically reported ICC links and label the related characteristics, which might introduce bias.
To mitigate this risk, 18 tag inference checkers are designed for double-checking.
For ICCs that can be triggered by multiple paths on the sliced code, we only record and label one path, which may influence the evaluation based on these labels.
Although this type of bias is difficult to avoid, we try to cover more ICCs to reduce the accidental errors brought by it.

%% file: content/conclusion.tex
\section{Conclusion}
Identifying ICC links precisely is essential to the analysis of apps.
However, the comprehensive evaluation of Android ICC resolution techniques faces big challenges due to the lack of high-quality datasets and metrics.
In this paper, we present multiple-type benchmark suites and design corresponding evaluation metrics.
For the oracle construction on real-world apps, we propose a dynamic ICC extraction approach and combine an automatic result filter and careful manual code auditing.
With both the constructed oracle set and the proposed metrics, we identified 38\%-85\% ICCs that are missed by tools and observed many wrongly reported ICCs.
Finally, based on the labeled characteristic tags, we discover the pros and cons of the state-of-the-art tools and summarize eight common FN/FP patterns for further improvement.
\section*{Acknowledgements}
The authors would like to thank the anonymous reviewers for their helpful comments and suggestions.
This work is supported by the National Natural Science Foundation of China (Grant No. 62102405 and Grant No. 62132020), the Key Research Program of Frontier Sciences, Chinese Academy of Sciences (Grant No. QYZDJ-SSW-JSC036), the Guangdong Basic and Applied Basic Research Fund (Grant No. 2021A1515011562), and the Guangdong Provincial Key Laboratory (Grant No. 2020B121201001).

%% file: sample-sigconf.bbl

\begin{thebibliography}{58}


\ifx \showCODEN    \undefined \def \showCODEN     #1{\unskip}     \fi
\ifx \showDOI      \undefined \def \showDOI       #1{#1}\fi
\ifx \showISBNx    \undefined \def \showISBNx     #1{\unskip}     \fi
\ifx \showISBNxiii \undefined \def \showISBNxiii  #1{\unskip}     \fi
\ifx \showISSN     \undefined \def \showISSN      #1{\unskip}     \fi
\ifx \showLCCN     \undefined \def \showLCCN      #1{\unskip}     \fi
\ifx \shownote     \undefined \def \shownote      #1{#1}          \fi
\ifx \showarticletitle \undefined \def \showarticletitle #1{#1}   \fi
\ifx \showURL      \undefined \def \showURL       {\relax}        \fi
\providecommand\bibfield[2]{#2}
\providecommand\bibinfo[2]{#2}
\providecommand\natexlab[1]{#1}
\providecommand\showeprint[2][]{arXiv:#2}

\bibitem[\protect\citeauthoryear{A3E}{A3E}{2016}]%
        {a3e}
\bibfield{author}{\bibinfo{person}{A3E}.} \bibinfo{year}{2016}\natexlab{}.
\newblock \bibinfo{title}{A3E}.
\newblock \bibinfo{howpublished}{\url{https://github.com/tanzirul/a3e}}.
\newblock


\bibitem[\protect\citeauthoryear{Ahmad, K{\"{a}}stner, Sunshine, and
  Aldrich}{Ahmad et~al\mbox{.}}{2016}]%
        {DBLP:conf/msr/AhmadKSA16}
\bibfield{author}{\bibinfo{person}{Waqar Ahmad}, \bibinfo{person}{Christian
  K{\"{a}}stner}, \bibinfo{person}{Joshua Sunshine}, {and}
  \bibinfo{person}{Jonathan Aldrich}.} \bibinfo{year}{2016}\natexlab{}.
\newblock \showarticletitle{Inter-app communication in Android: developer
  challenges}. In \bibinfo{booktitle}{\emph{Proceedings of the 13th
  International Conference on Mining Software Repositories, {MSR} 2016, Austin,
  TX, USA, May 14-22, 2016}}, \bibfield{editor}{\bibinfo{person}{Miryung Kim},
  \bibinfo{person}{Romain Robbes}, {and} \bibinfo{person}{Christian Bird}}
  (Eds.). \bibinfo{publisher}{{ACM}}, \bibinfo{pages}{177--188}.
\newblock


\bibitem[\protect\citeauthoryear{Allix, Bissyand{\'e}, Klein, and
  Le~Traon}{Allix et~al\mbox{.}}{2016}]%
        {Allix:2016:ACM:2901739.2903508}
\bibfield{author}{\bibinfo{person}{Kevin Allix},
  \bibinfo{person}{Tegawend{\'e}~F. Bissyand{\'e}}, \bibinfo{person}{Jacques
  Klein}, {and} \bibinfo{person}{Yves Le~Traon}.}
  \bibinfo{year}{2016}\natexlab{}.
\newblock \showarticletitle{AndroZoo: Collecting Millions of Android Apps for
  the Research Community}. In \bibinfo{booktitle}{\emph{Proceedings of the 13th
  International Conference on Mining Software Repositories}} (Austin, Texas)
  \emph{(\bibinfo{series}{MSR '16})}. \bibinfo{publisher}{ACM},
  \bibinfo{pages}{468--471}.
\newblock
\showISBNx{978-1-4503-4186-8}


\bibitem[\protect\citeauthoryear{Arzt, Rasthofer, Fritz, Bodden, Bartel, Klein,
  Traon, Octeau, and McDaniel}{Arzt et~al\mbox{.}}{2014}]%
        {Arzt14PLDI}
\bibfield{author}{\bibinfo{person}{Steven Arzt}, \bibinfo{person}{Siegfried
  Rasthofer}, \bibinfo{person}{Christian Fritz}, \bibinfo{person}{Eric Bodden},
  \bibinfo{person}{Alexandre Bartel}, \bibinfo{person}{Jacques Klein},
  \bibinfo{person}{Yves~Le Traon}, \bibinfo{person}{Damien Octeau}, {and}
  \bibinfo{person}{Patrick McDaniel}.} \bibinfo{year}{2014}\natexlab{}.
\newblock \showarticletitle{{FlowDroid}: precise context, flow, field,
  object-sensitive and lifecycle-aware taint analysis for {Android} apps}. In
  \bibinfo{booktitle}{\emph{{PLDI} 2014}}. \bibinfo{pages}{29}.
\newblock


\bibitem[\protect\citeauthoryear{Azim and Neamtiu}{Azim and Neamtiu}{2013}]%
        {DBLP:conf/oopsla/AzimN13}
\bibfield{author}{\bibinfo{person}{Tanzirul Azim} {and} \bibinfo{person}{Iulian
  Neamtiu}.} \bibinfo{year}{2013}\natexlab{}.
\newblock \showarticletitle{Targeted and depth-first exploration for systematic
  testing of {Android} apps}. In \bibinfo{booktitle}{\emph{{OOPSLA} 2013, part
  of {SPLASH} 2013}}. \bibinfo{pages}{641--660}.
\newblock


\bibitem[\protect\citeauthoryear{Bagheri, Sadeghi, Garcia, and Malek}{Bagheri
  et~al\mbox{.}}{2015}]%
        {DBLP:journals/tse/BagheriSGM15}
\bibfield{author}{\bibinfo{person}{Hamid Bagheri}, \bibinfo{person}{Alireza
  Sadeghi}, \bibinfo{person}{Joshua Garcia}, {and} \bibinfo{person}{Sam
  Malek}.} \bibinfo{year}{2015}\natexlab{}.
\newblock \showarticletitle{{COVERT:} Compositional Analysis of Android
  Inter-App Permission Leakage}.
\newblock \bibinfo{journal}{\emph{TSE}} \bibinfo{volume}{41},
  \bibinfo{number}{9} (\bibinfo{year}{2015}), \bibinfo{pages}{866--886}.
\newblock


\bibitem[\protect\citeauthoryear{Bhandari, Herbreteau, Laxmi, Zemmari, Gaur,
  and Roop}{Bhandari et~al\mbox{.}}{2020}]%
        {DBLP:journals/fgcs/BhandariHLZGR20}
\bibfield{author}{\bibinfo{person}{Shweta Bhandari},
  \bibinfo{person}{Fr{\'{e}}d{\'{e}}ric Herbreteau}, \bibinfo{person}{Vijay
  Laxmi}, \bibinfo{person}{Akka Zemmari}, \bibinfo{person}{Manoj~Singh Gaur},
  {and} \bibinfo{person}{Partha~S. Roop}.} \bibinfo{year}{2020}\natexlab{}.
\newblock \showarticletitle{\emph{SneakLeak+}: Large-scale klepto apps
  analysis}.
\newblock \bibinfo{journal}{\emph{Future Gener. Comput. Syst.}}
  \bibinfo{volume}{109} (\bibinfo{year}{2020}), \bibinfo{pages}{593--603}.
\newblock


\bibitem[\protect\citeauthoryear{Bhandari, Jaballah, Jain, Laxmi, Zemmari,
  Gaur, Mosbah, and Conti}{Bhandari et~al\mbox{.}}{2017}]%
        {DBLP:journals/compsec/BhandariJJLZGMC17}
\bibfield{author}{\bibinfo{person}{Shweta Bhandari}, \bibinfo{person}{Wafa~Ben
  Jaballah}, \bibinfo{person}{Vineeta Jain}, \bibinfo{person}{Vijay Laxmi},
  \bibinfo{person}{Akka Zemmari}, \bibinfo{person}{Manoj~Singh Gaur},
  \bibinfo{person}{Mohamed Mosbah}, {and} \bibinfo{person}{Mauro Conti}.}
  \bibinfo{year}{2017}\natexlab{}.
\newblock \showarticletitle{Android inter-app communication threats and
  detection techniques}.
\newblock \bibinfo{journal}{\emph{Comput. Secur.}}  \bibinfo{volume}{70}
  (\bibinfo{year}{2017}), \bibinfo{pages}{392--421}.
\newblock


\bibitem[\protect\citeauthoryear{Bohluli and Shahriari}{Bohluli and
  Shahriari}{2018}]%
        {DBLP:conf/iscisc/BohluliS18}
\bibfield{author}{\bibinfo{person}{Zohreh Bohluli} {and}
  \bibinfo{person}{Hamid~Reza Shahriari}.} \bibinfo{year}{2018}\natexlab{}.
\newblock \showarticletitle{Detecting Privacy Leaks in Android Apps using
  Inter-Component Information Flow Control Analysis}. In
  \bibinfo{booktitle}{\emph{15th International {ISC} (Iranian Society of
  Cryptology) Conference on Information Security and Cryptology, {ISCISC} 2018,
  Tehran, Iran, August 28-29, 2018}}. \bibinfo{publisher}{{IEEE}},
  \bibinfo{pages}{1--6}.
\newblock


\bibitem[\protect\citeauthoryear{Bosu, Liu, Yao, and Wang}{Bosu
  et~al\mbox{.}}{2017}]%
        {DBLP:conf/ccs/BosuLYW17}
\bibfield{author}{\bibinfo{person}{Amiangshu Bosu}, \bibinfo{person}{Fang Liu},
  \bibinfo{person}{Danfeng~(Daphne) Yao}, {and} \bibinfo{person}{Gang Wang}.}
  \bibinfo{year}{2017}\natexlab{}.
\newblock \showarticletitle{Collusive Data Leak and More: Large-scale Threat
  Analysis of Inter-app Communications}. In
  \bibinfo{booktitle}{\emph{Proceedings of the 2017 {ACM} on Asia Conference on
  Computer and Communications Security, AsiaCCS 2017, Abu Dhabi, United Arab
  Emirates, April 2-6, 2017}}. \bibinfo{publisher}{{ACM}},
  \bibinfo{pages}{71--85}.
\newblock


\bibitem[\protect\citeauthoryear{Chang, Lei, Sun, Wang, Jing, He, and
  Wang}{Chang et~al\mbox{.}}{2021}]%
        {DBLP:journals/tdsc/ChangLSWJHW21}
\bibfield{author}{\bibinfo{person}{Huan Chang}, \bibinfo{person}{Lingguang
  Lei}, \bibinfo{person}{Kun Sun}, \bibinfo{person}{Yuewu Wang},
  \bibinfo{person}{Jiwu Jing}, \bibinfo{person}{Yi He}, {and}
  \bibinfo{person}{Pingjian Wang}.} \bibinfo{year}{2021}\natexlab{}.
\newblock \showarticletitle{Vulnerable Service Invocation and Countermeasures}.
\newblock \bibinfo{journal}{\emph{{IEEE} Trans. Dependable Secur. Comput.}}
  \bibinfo{volume}{18}, \bibinfo{number}{4} (\bibinfo{year}{2021}),
  \bibinfo{pages}{1733--1750}.
\newblock


\bibitem[\protect\citeauthoryear{Chen, Fan, Chen, and Liu}{Chen
  et~al\mbox{.}}{2022}]%
        {chen2019storydistiller}
\bibfield{author}{\bibinfo{person}{Sen Chen}, \bibinfo{person}{Lingling Fan},
  \bibinfo{person}{Chunyang Chen}, {and} \bibinfo{person}{Yang Liu}.}
  \bibinfo{year}{2022}\natexlab{}.
\newblock \showarticletitle{Automatically Distilling Storyboard with Rich
  Features for Android Apps}. In \bibinfo{booktitle}{\emph{IEEE Transactions on
  Software Engineering (TSE)}}. IEEE.
\newblock


\bibitem[\protect\citeauthoryear{Chen, Fan, Chen, Su, Li, Liu, and Xu}{Chen
  et~al\mbox{.}}{2019}]%
        {DBLP:conf/icse/ChenFCSLLX19}
\bibfield{author}{\bibinfo{person}{Sen Chen}, \bibinfo{person}{Lingling Fan},
  \bibinfo{person}{Chunyang Chen}, \bibinfo{person}{Ting Su},
  \bibinfo{person}{Wenhe Li}, \bibinfo{person}{Yang Liu}, {and}
  \bibinfo{person}{Lihua Xu}.} \bibinfo{year}{2019}\natexlab{}.
\newblock \showarticletitle{StoryDroid: automated generation of storyboard for
  Android apps}. In \bibinfo{booktitle}{\emph{Proceedings of the 41st
  International Conference on Software Engineering, {ICSE} 2019, Montreal, QC,
  Canada, May 25-31, 2019}}. \bibinfo{publisher}{{IEEE} / {ACM}},
  \bibinfo{pages}{596--607}.
\newblock


\bibitem[\protect\citeauthoryear{Cheng, Shi, Gong, and Guan}{Cheng
  et~al\mbox{.}}{2021}]%
        {DBLP:journals/di/ChengSGG21}
\bibfield{author}{\bibinfo{person}{Chris~Chao{-}Chun Cheng},
  \bibinfo{person}{Chen Shi}, \bibinfo{person}{Neil~Zhenqiang Gong}, {and}
  \bibinfo{person}{Yong Guan}.} \bibinfo{year}{2021}\natexlab{}.
\newblock \showarticletitle{LogExtractor: Extracting digital evidence from
  android log messages via string and taint analysis}.
\newblock \bibinfo{journal}{\emph{Digit. Investig.}}  \bibinfo{volume}{37
  Supplement} (\bibinfo{year}{2021}), \bibinfo{pages}{301193}.
\newblock


\bibitem[\protect\citeauthoryear{Component}{Component}{2022}]%
        {component}
\bibfield{author}{\bibinfo{person}{Component}.}
  \bibinfo{year}{2022}\natexlab{}.
\newblock \bibinfo{title}{Component}.
\newblock
  \bibinfo{howpublished}{\url{https://developer.android.com/guide/components/fundamentals\#Components}}.
\newblock


\bibitem[\protect\citeauthoryear{DroidBench}{DroidBench}{2017}]%
        {DroidBench}
\bibfield{author}{\bibinfo{person}{DroidBench}.}
  \bibinfo{year}{2017}\natexlab{}.
\newblock \bibinfo{title}{DroidBench}.
\newblock
  \bibinfo{howpublished}{\url{https://github.com/secure-software-engineering/DroidBench}}.
\newblock


\bibitem[\protect\citeauthoryear{Elish, Cai, Barton, Yao, and Ryder}{Elish
  et~al\mbox{.}}{2020}]%
        {DBLP:journals/tmc/ElishCBYR20}
\bibfield{author}{\bibinfo{person}{Karim~O. Elish}, \bibinfo{person}{Haipeng
  Cai}, \bibinfo{person}{Daniel Barton}, \bibinfo{person}{Danfeng Yao}, {and}
  \bibinfo{person}{Barbara~G. Ryder}.} \bibinfo{year}{2020}\natexlab{}.
\newblock \showarticletitle{Identifying Mobile Inter-App Communication Risks}.
\newblock \bibinfo{journal}{\emph{{IEEE} Trans. Mob. Comput.}}
  \bibinfo{volume}{19}, \bibinfo{number}{1} (\bibinfo{year}{2020}),
  \bibinfo{pages}{90--102}.
\newblock


\bibitem[\protect\citeauthoryear{{F-Droid}}{{F-Droid}}{2019}]%
        {F-Droid}
\bibfield{author}{\bibinfo{person}{{F-Droid}}.}
  \bibinfo{year}{2019}\natexlab{}.
\newblock \bibinfo{howpublished}{\url{https://f-droid.org/}}.
\newblock


\bibitem[\protect\citeauthoryear{Fan, Su, Chen, Meng, Liu, Xu, and Pu}{Fan
  et~al\mbox{.}}{2018}]%
        {DBLP:conf/kbse/FanSCMLXP18}
\bibfield{author}{\bibinfo{person}{Lingling Fan}, \bibinfo{person}{Ting Su},
  \bibinfo{person}{Sen Chen}, \bibinfo{person}{Guozhu Meng},
  \bibinfo{person}{Yang Liu}, \bibinfo{person}{Lihua Xu}, {and}
  \bibinfo{person}{Geguang Pu}.} \bibinfo{year}{2018}\natexlab{}.
\newblock \showarticletitle{Efficiently manifesting asynchronous programming
  errors in Android apps}. In \bibinfo{booktitle}{\emph{Proceedings of the 33rd
  {ACM/IEEE} International Conference on Automated Software Engineering, {ASE}
  2018, Montpellier, France, September 3-7, 2018}}. \bibinfo{publisher}{{ACM}},
  \bibinfo{pages}{486--497}.
\newblock


\bibitem[\protect\citeauthoryear{Fragment}{Fragment}{2022}]%
        {fragment}
\bibfield{author}{\bibinfo{person}{Fragment}.} \bibinfo{year}{2022}\natexlab{}.
\newblock \bibinfo{title}{Fragment}.
\newblock
  \bibinfo{howpublished}{\url{https://developer.android.com/guide/fragments}}.
\newblock


\bibitem[\protect\citeauthoryear{GATOR}{GATOR}{2019}]%
        {GATOR}
\bibfield{author}{\bibinfo{person}{GATOR}.} \bibinfo{year}{2019}\natexlab{}.
\newblock \bibinfo{title}{GATOR}.
\newblock
  \bibinfo{howpublished}{\url{http://web.cse.ohio-state.edu/presto/software/gator/}}.
\newblock


\bibitem[\protect\citeauthoryear{Gordon, Kim, Perkins, Gilham, Nguyen, and
  Rinard}{Gordon et~al\mbox{.}}{2015}]%
        {DBLP:conf/ndss/GordonKPGNR15}
\bibfield{author}{\bibinfo{person}{Michael~I. Gordon},
  \bibinfo{person}{Deokhwan Kim}, \bibinfo{person}{Jeff~H. Perkins},
  \bibinfo{person}{Limei Gilham}, \bibinfo{person}{Nguyen Nguyen}, {and}
  \bibinfo{person}{Martin~C. Rinard}.} \bibinfo{year}{2015}\natexlab{}.
\newblock \showarticletitle{Information Flow Analysis of Android Applications
  in DroidSafe}. In \bibinfo{booktitle}{\emph{22nd Annual Network and
  Distributed System Security Symposium, {NDSS} 2015, San Diego, California,
  USA, February 8-11, 2015}}. \bibinfo{publisher}{The Internet Society}.
\newblock


\bibitem[\protect\citeauthoryear{Gu, Sun, Ma, Cao, Xu, Yao, Zhang, Lu, and
  Su}{Gu et~al\mbox{.}}{2019}]%
        {DBLP:conf/icse/GuSMC0YZLS19}
\bibfield{author}{\bibinfo{person}{Tianxiao Gu}, \bibinfo{person}{Chengnian
  Sun}, \bibinfo{person}{Xiaoxing Ma}, \bibinfo{person}{Chun Cao},
  \bibinfo{person}{Chang Xu}, \bibinfo{person}{Yuan Yao},
  \bibinfo{person}{Qirun Zhang}, \bibinfo{person}{Jian Lu}, {and}
  \bibinfo{person}{Zhendong Su}.} \bibinfo{year}{2019}\natexlab{}.
\newblock \showarticletitle{Practical {GUI} testing of Android applications via
  model abstraction and refinement}. In \bibinfo{booktitle}{\emph{Proceedings
  of the 41st International Conference on Software Engineering, {ICSE} 2019,
  Montreal, QC, Canada, May 25-31, 2019}}. \bibinfo{pages}{269--280}.
\newblock


\bibitem[\protect\citeauthoryear{IC3}{IC3}{2015}]%
        {IC3}
\bibfield{author}{\bibinfo{person}{IC3}.} \bibinfo{year}{2015}\natexlab{}.
\newblock \bibinfo{title}{IC3}.
\newblock \bibinfo{howpublished}{\url{https://github.com/siis/ic3}}.
\newblock


\bibitem[\protect\citeauthoryear{IC3-DIALDroid}{IC3-DIALDroid}{2020}]%
        {IC3-DIALDroid}
\bibfield{author}{\bibinfo{person}{IC3-DIALDroid}.}
  \bibinfo{year}{2020}\natexlab{}.
\newblock \bibinfo{title}{IC3-DIALDroid}.
\newblock
  \bibinfo{howpublished}{\url{https://github.com/dialdroid-android/ic3-dialdroid}}.
\newblock


\bibitem[\protect\citeauthoryear{ICC-Bench}{ICC-Bench}{2017}]%
        {ICC-Bench}
\bibfield{author}{\bibinfo{person}{ICC-Bench}.}
  \bibinfo{year}{2017}\natexlab{}.
\newblock \bibinfo{title}{ICC-Bench}.
\newblock \bibinfo{howpublished}{\url{https://github.com/fgwei/ICC-Bench}}.
\newblock


\bibitem[\protect\citeauthoryear{ICC-Resolution-Evaluation}{ICC-Resolution-Evaluation}{2022}]%
        {ICC-Resolution-Evaluation}
\bibfield{author}{\bibinfo{person}{ICC-Resolution-Evaluation}.}
  \bibinfo{year}{2022}\natexlab{}.
\newblock \bibinfo{title}{ICC-Resolution-Evaluation}.
\newblock
  \bibinfo{howpublished}{\url{https://github.com/hanada31/ICC-Resolution-Evaluation/}}.
\newblock


\bibitem[\protect\citeauthoryear{ICCBot}{ICCBot}{2022}]%
        {ICCBot}
\bibfield{author}{\bibinfo{person}{ICCBot}.} \bibinfo{year}{2022}\natexlab{}.
\newblock \bibinfo{title}{ICCBot}.
\newblock \bibinfo{howpublished}{\url{https://github.com/hanada31/ICCBot}}.
\newblock


\bibitem[\protect\citeauthoryear{ICCViewer}{ICCViewer}{2022}]%
        {ICCViewer}
\bibfield{author}{\bibinfo{person}{ICCViewer}.}
  \bibinfo{year}{2022}\natexlab{}.
\newblock \bibinfo{title}{ICCViewer}.
\newblock \bibinfo{howpublished}{\url{https://iccviewer.ldby.site/ICCViewer/}}.
\newblock


\bibitem[\protect\citeauthoryear{Intent}{Intent}{2022}]%
        {Intent}
\bibfield{author}{\bibinfo{person}{Intent}.} \bibinfo{year}{2022}\natexlab{}.
\newblock \bibinfo{title}{Intent}.
\newblock
  \bibinfo{howpublished}{\url{https://developer.android.com/guide/components/intents-filters}}.
\newblock


\bibitem[\protect\citeauthoryear{Jabbarvand, Lin, and Malek}{Jabbarvand
  et~al\mbox{.}}{2019}]%
        {DBLP:conf/icse/JabbarvandLM19}
\bibfield{author}{\bibinfo{person}{Reyhaneh Jabbarvand},
  \bibinfo{person}{Jun{-}Wei Lin}, {and} \bibinfo{person}{Sam Malek}.}
  \bibinfo{year}{2019}\natexlab{}.
\newblock \showarticletitle{Search-based energy testing of Android}. In
  \bibinfo{booktitle}{\emph{Proceedings of the 41st International Conference on
  Software Engineering, {ICSE} 2019, Montreal, QC, Canada, May 25-31, 2019}}.
  \bibinfo{publisher}{{IEEE} / {ACM}}, \bibinfo{pages}{1119--1130}.
\newblock


\bibitem[\protect\citeauthoryear{Jha and Lee}{Jha and Lee}{[n.d.]}]%
        {jhaiccmatt}
\bibfield{author}{\bibinfo{person}{Ajay~Kumar Jha} {and}
  \bibinfo{person}{Woo~Jin Lee}.} \bibinfo{year}{[n.d.]}\natexlab{}.
\newblock \showarticletitle{ICCMATT: Modeling, analysis, and test case
  generation of inter-component communication in Android}.
\newblock  (\bibinfo{year}{[n.\,d.]}).
\newblock


\bibitem[\protect\citeauthoryear{Kim, Yoon, Yi, and Shin}{Kim
  et~al\mbox{.}}{2012}]%
        {SCANDAL}
\bibfield{author}{\bibinfo{person}{Jinyung Kim}, \bibinfo{person}{Yongho Yoon},
  \bibinfo{person}{Kwangkeun Yi}, {and} \bibinfo{person}{Junbum Shin}.}
  \bibinfo{year}{2012}\natexlab{}.
\newblock \showarticletitle{SCANDAL: Static Analyzer for Detecting Privacy
  Leaks in Android Applications}.
\newblock \bibinfo{journal}{\emph{{MoST}}} \bibinfo{volume}{12},
  \bibinfo{number}{110} (\bibinfo{year}{2012}), \bibinfo{pages}{1}.
\newblock


\bibitem[\protect\citeauthoryear{Lai and Rubin}{Lai and Rubin}{2019}]%
        {DBLP:conf/kbse/LaiR19}
\bibfield{author}{\bibinfo{person}{Duling Lai} {and} \bibinfo{person}{Julia
  Rubin}.} \bibinfo{year}{2019}\natexlab{}.
\newblock \showarticletitle{Goal-Driven Exploration for Android Applications}.
  In \bibinfo{booktitle}{\emph{34th {IEEE/ACM} International Conference on
  Automated Software Engineering, {ASE} 2019, San Diego, CA, USA, November
  11-15, 2019}}. \bibinfo{publisher}{{IEEE}}, \bibinfo{pages}{115--127}.
\newblock


\bibitem[\protect\citeauthoryear{Lee, Bang, Safi, Shahbazian, Zhao, and
  Medvidovic}{Lee et~al\mbox{.}}{2017}]%
        {DBLP:conf/icse/LeeBSSZM17}
\bibfield{author}{\bibinfo{person}{Youn~Kyu Lee}, \bibinfo{person}{Jae~Young
  Bang}, \bibinfo{person}{Gholamreza Safi}, \bibinfo{person}{Arman Shahbazian},
  \bibinfo{person}{Yixue Zhao}, {and} \bibinfo{person}{Nenad Medvidovic}.}
  \bibinfo{year}{2017}\natexlab{}.
\newblock \showarticletitle{A \emph{SEALANT} for inter-app security holes in
  android}. In \bibinfo{booktitle}{\emph{Proceedings of the 39th International
  Conference on Software Engineering, {ICSE} 2017, Buenos Aires, Argentina, May
  20-28, 2017}}. \bibinfo{publisher}{{IEEE} / {ACM}},
  \bibinfo{pages}{312--323}.
\newblock


\bibitem[\protect\citeauthoryear{Li, Bartel, Bissyand{\'{e}}, Klein, Traon,
  Arzt, Rasthofer, Bodden, Octeau, and McDaniel}{Li et~al\mbox{.}}{2015}]%
        {Li15ICSE}
\bibfield{author}{\bibinfo{person}{Li Li}, \bibinfo{person}{Alexandre Bartel},
  \bibinfo{person}{Tegawend{\'{e}}~F. Bissyand{\'{e}}},
  \bibinfo{person}{Jacques Klein}, \bibinfo{person}{Yves~Le Traon},
  \bibinfo{person}{Steven Arzt}, \bibinfo{person}{Siegfried Rasthofer},
  \bibinfo{person}{Eric Bodden}, \bibinfo{person}{Damien Octeau}, {and}
  \bibinfo{person}{Patrick McDaniel}.} \bibinfo{year}{2015}\natexlab{}.
\newblock \showarticletitle{{IccTA}: Detecting Inter-Component Privacy Leaks in
  {Android} Apps}. In \bibinfo{booktitle}{\emph{{ICSE} 2015}}.
  \bibinfo{pages}{280--291}.
\newblock


\bibitem[\protect\citeauthoryear{Li, Bissyand{\'{e}}, Papadakis, Rasthofer,
  Bartel, Octeau, Klein, and Traon}{Li et~al\mbox{.}}{2017}]%
        {DBLP:journals/infsof/LiBPRBOKT17}
\bibfield{author}{\bibinfo{person}{Li Li}, \bibinfo{person}{Tegawend{\'{e}}~F.
  Bissyand{\'{e}}}, \bibinfo{person}{Mike Papadakis},
  \bibinfo{person}{Siegfried Rasthofer}, \bibinfo{person}{Alexandre Bartel},
  \bibinfo{person}{Damien Octeau}, \bibinfo{person}{Jacques Klein}, {and}
  \bibinfo{person}{Yves~Le Traon}.} \bibinfo{year}{2017}\natexlab{}.
\newblock \showarticletitle{Static analysis of android apps: {A} systematic
  literature review}.
\newblock \bibinfo{journal}{\emph{Inf. Softw. Technol.}}  \bibinfo{volume}{88}
  (\bibinfo{year}{2017}), \bibinfo{pages}{67--95}.
\newblock


\bibitem[\protect\citeauthoryear{Lu, Li, Wu, Lee, and Jiang}{Lu
  et~al\mbox{.}}{2012}]%
        {DBLP:conf/ccs/LuLWLJ12}
\bibfield{author}{\bibinfo{person}{Long Lu}, \bibinfo{person}{Zhichun Li},
  \bibinfo{person}{Zhenyu Wu}, \bibinfo{person}{Wenke Lee}, {and}
  \bibinfo{person}{Guofei Jiang}.} \bibinfo{year}{2012}\natexlab{}.
\newblock \showarticletitle{{CHEX:} statically vetting Android apps for
  component hijacking vulnerabilities}. In \bibinfo{booktitle}{\emph{the {ACM}
  Conference on Computer and Communications Security, CCS'12, Raleigh, NC, USA,
  October 16-18, 2012}}. \bibinfo{publisher}{{ACM}}, \bibinfo{pages}{229--240}.
\newblock


\bibitem[\protect\citeauthoryear{Octeau, Luchaup, Dering, Jha, and
  McDaniel}{Octeau et~al\mbox{.}}{2015}]%
        {Octeau15ICSE}
\bibfield{author}{\bibinfo{person}{Damien Octeau}, \bibinfo{person}{Daniel
  Luchaup}, \bibinfo{person}{Matthew Dering}, \bibinfo{person}{Somesh Jha},
  {and} \bibinfo{person}{Patrick McDaniel}.} \bibinfo{year}{2015}\natexlab{}.
\newblock \showarticletitle{{Composite Constant Propagation: Application to
  {Android} Inter-Component Communication Analysis}}. In
  \bibinfo{booktitle}{\emph{{ICSE} 2015}}. \bibinfo{pages}{77--88}.
\newblock


\bibitem[\protect\citeauthoryear{Octeau, Luchaup, Jha, and McDaniel}{Octeau
  et~al\mbox{.}}{2016}]%
        {DBLP:journals/tse/OcteauLJM16}
\bibfield{author}{\bibinfo{person}{Damien Octeau}, \bibinfo{person}{Daniel
  Luchaup}, \bibinfo{person}{Somesh Jha}, {and} \bibinfo{person}{Patrick~D.
  McDaniel}.} \bibinfo{year}{2016}\natexlab{}.
\newblock \showarticletitle{Composite Constant Propagation and its Application
  to Android Program Analysis}.
\newblock \bibinfo{journal}{\emph{{IEEE} Trans. Software Eng.}}
  \bibinfo{volume}{42}, \bibinfo{number}{11} (\bibinfo{year}{2016}),
  \bibinfo{pages}{999--1014}.
\newblock


\bibitem[\protect\citeauthoryear{Octeau, McDaniel, Jha, Bartel, Bodden, Klein,
  and Traon}{Octeau et~al\mbox{.}}{2013}]%
        {DBLP:conf/uss/OcteauMJBBKT13}
\bibfield{author}{\bibinfo{person}{Damien Octeau}, \bibinfo{person}{Patrick~D.
  McDaniel}, \bibinfo{person}{Somesh Jha}, \bibinfo{person}{Alexandre Bartel},
  \bibinfo{person}{Eric Bodden}, \bibinfo{person}{Jacques Klein}, {and}
  \bibinfo{person}{Yves~Le Traon}.} \bibinfo{year}{2013}\natexlab{}.
\newblock \showarticletitle{Effective Inter-Component Communication Mapping in
  Android: An Essential Step Towards Holistic Security Analysis}. In
  \bibinfo{booktitle}{\emph{Proceedings of the 22th {USENIX} Security
  Symposium, Washington, DC, USA, August 14-16, 2013}}.
  \bibinfo{publisher}{{USENIX} Association}, \bibinfo{pages}{543--558}.
\newblock


\bibitem[\protect\citeauthoryear{Play}{Play}{2019}]%
        {GooglePlay}
\bibfield{author}{\bibinfo{person}{Google Play}.}
  \bibinfo{year}{2019}\natexlab{}.
\newblock \bibinfo{howpublished}{\url{https://play.google.com/store}}.
\newblock


\bibitem[\protect\citeauthoryear{RAICC-Bench}{RAICC-Bench}{2021}]%
        {RAICC-Bench}
\bibfield{author}{\bibinfo{person}{RAICC-Bench}.}
  \bibinfo{year}{2021}\natexlab{}.
\newblock \bibinfo{title}{RAICC-Bench}.
\newblock
  \bibinfo{howpublished}{\url{https://github.com/Trustworthy-Software/RAICC}}.
\newblock


\bibitem[\protect\citeauthoryear{Sadeghi, Jabbarvand, Ghorbani, Bagheri, and
  Malek}{Sadeghi et~al\mbox{.}}{[n.d.]}]%
        {DBLP:conf/icse/SadeghiJGBM18}
\bibfield{author}{\bibinfo{person}{Alireza Sadeghi}, \bibinfo{person}{Reyhaneh
  Jabbarvand}, \bibinfo{person}{Negar Ghorbani}, \bibinfo{person}{Hamid
  Bagheri}, {and} \bibinfo{person}{Sam Malek}.}
  \bibinfo{year}{[n.d.]}\natexlab{}.
\newblock \showarticletitle{A temporal permission analysis and enforcement
  framework for Android}. In \bibinfo{booktitle}{\emph{Proceedings of the 40th
  International Conference on Software Engineering, {ICSE} 2018, Gothenburg,
  Sweden, May 27 - June 03, 2018}}, \bibfield{editor}{\bibinfo{person}{Michel
  Chaudron}, \bibinfo{person}{Ivica Crnkovic}, \bibinfo{person}{Marsha
  Chechik}, {and} \bibinfo{person}{Mark Harman}} (Eds.).
  \bibinfo{pages}{846--857}.
\newblock


\bibitem[\protect\citeauthoryear{Samhi, Bartel, Bissyand{\'{e}}, and
  Klein}{Samhi et~al\mbox{.}}{2021}]%
        {DBLP:conf/icse/SamhiBBK21}
\bibfield{author}{\bibinfo{person}{Jordan Samhi}, \bibinfo{person}{Alexandre
  Bartel}, \bibinfo{person}{Tegawend{\'{e}}~F. Bissyand{\'{e}}}, {and}
  \bibinfo{person}{Jacques Klein}.} \bibinfo{year}{2021}\natexlab{}.
\newblock \showarticletitle{{RAICC:} Revealing Atypical Inter-Component
  Communication in Android Apps}. In \bibinfo{booktitle}{\emph{43rd {IEEE/ACM}
  International Conference on Software Engineering, {ICSE} 2021, Madrid, Spain,
  22-30 May 2021}}. \bibinfo{publisher}{{IEEE}}, \bibinfo{pages}{1398--1409}.
\newblock


\bibitem[\protect\citeauthoryear{StoryDistiller}{StoryDistiller}{2022}]%
        {StoryDistiller}
\bibfield{author}{\bibinfo{person}{StoryDistiller}.}
  \bibinfo{year}{2022}\natexlab{}.
\newblock \bibinfo{title}{StoryDistiller}.
\newblock
  \bibinfo{howpublished}{\url{https://github.com/tjusenchen/StoryDistiller}}.
\newblock


\bibitem[\protect\citeauthoryear{StoryDroid-Bench}{StoryDroid-Bench}{2019}]%
        {StoryDroid-Bench}
\bibfield{author}{\bibinfo{person}{StoryDroid-Bench}.}
  \bibinfo{year}{2019}\natexlab{}.
\newblock \bibinfo{title}{StoryDroid-Bench}.
\newblock
  \bibinfo{howpublished}{\url{https://sites.google.com/view/storydroid/}}.
\newblock


\bibitem[\protect\citeauthoryear{Tang, Sui, Wang, Luo, Zhou, and Xu}{Tang
  et~al\mbox{.}}{[n.d.]}]%
        {DBLP:conf/sigsoft/TangSWLZ020}
\bibfield{author}{\bibinfo{person}{Yutian Tang}, \bibinfo{person}{Yulei Sui},
  \bibinfo{person}{Haoyu Wang}, \bibinfo{person}{Xiapu Luo},
  \bibinfo{person}{Hao Zhou}, {and} \bibinfo{person}{Zhou Xu}.}
  \bibinfo{year}{[n.d.]}\natexlab{}.
\newblock \showarticletitle{All your app links are belong to us: understanding
  the threats of instant apps based attacks}. In
  \bibinfo{booktitle}{\emph{{ESEC/FSE} '20: 28th {ACM} Joint European Software
  Engineering Conference and Symposium on the Foundations of Software
  Engineering, Virtual Event, USA, November 8-13, 2020}},
  \bibfield{editor}{\bibinfo{person}{Prem Devanbu}, \bibinfo{person}{Myra~B.
  Cohen}, {and} \bibinfo{person}{Thomas Zimmermann}} (Eds.).
  \bibinfo{pages}{914--926}.
\newblock


\bibitem[\protect\citeauthoryear{Tsutano, Bachala, Srisa{-}an, Rothermel, and
  Dinh}{Tsutano et~al\mbox{.}}{2017}]%
        {DBLP:conf/icse/TsutanoBSRD17}
\bibfield{author}{\bibinfo{person}{Yutaka Tsutano}, \bibinfo{person}{Shakthi
  Bachala}, \bibinfo{person}{Witawas Srisa{-}an}, \bibinfo{person}{Gregg
  Rothermel}, {and} \bibinfo{person}{Jackson Dinh}.}
  \bibinfo{year}{2017}\natexlab{}.
\newblock \showarticletitle{An efficient, robust, and scalable approach for
  analyzing interacting android apps}. In \bibinfo{booktitle}{\emph{Proceedings
  of the 39th International Conference on Software Engineering, {ICSE} 2017,
  Buenos Aires, Argentina, May 20-28, 2017}},
  \bibfield{editor}{\bibinfo{person}{Sebasti{\'{a}}n Uchitel},
  \bibinfo{person}{Alessandro Orso}, {and} \bibinfo{person}{Martin~P.
  Robillard}} (Eds.). \bibinfo{publisher}{{IEEE} / {ACM}},
  \bibinfo{pages}{324--334}.
\newblock


\bibitem[\protect\citeauthoryear{Wei, Roy, Ou, and Robby}{Wei
  et~al\mbox{.}}{2014}]%
        {DBLP:conf/ccs/WeiROR14}
\bibfield{author}{\bibinfo{person}{Fengguo Wei}, \bibinfo{person}{Sankardas
  Roy}, \bibinfo{person}{Xinming Ou}, {and} \bibinfo{person}{Robby}.}
  \bibinfo{year}{2014}\natexlab{}.
\newblock \showarticletitle{Amandroid: {A} Precise and General Inter-component
  Data Flow Analysis Framework for Security Vetting of Android Apps}. In
  \bibinfo{booktitle}{\emph{Proceedings of the 2014 {ACM} {SIGSAC} Conference
  on Computer and Communications Security, Scottsdale, AZ, USA, November 3-7,
  2014}}. \bibinfo{publisher}{{ACM}}, \bibinfo{pages}{1329--1341}.
\newblock


\bibitem[\protect\citeauthoryear{Yan, Liu, Pan, Yan, Zhang, and Liang}{Yan
  et~al\mbox{.}}{2020}]%
        {DBLP:conf/icse/YanLP00L20}
\bibfield{author}{\bibinfo{person}{Jiwei Yan}, \bibinfo{person}{Hao Liu},
  \bibinfo{person}{Linjie Pan}, \bibinfo{person}{Jun Yan},
  \bibinfo{person}{Jian Zhang}, {and} \bibinfo{person}{Bin Liang}.}
  \bibinfo{year}{2020}\natexlab{}.
\newblock \showarticletitle{Multiple-entry testing of Android applications by
  constructing activity launching contexts}. In
  \bibinfo{booktitle}{\emph{{ICSE} '20: 42nd International Conference on
  Software Engineering, Seoul, South Korea, 27 June - 19 July, 2020}}.
  \bibinfo{pages}{457--468}.
\newblock


\bibitem[\protect\citeauthoryear{Yan, Zhang, Liu, Yan, and Zhang}{Yan
  et~al\mbox{.}}{2022}]%
        {icse22_iccbot}
\bibfield{author}{\bibinfo{person}{Jiwei Yan}, \bibinfo{person}{Shixin Zhang},
  \bibinfo{person}{Yepang Liu}, \bibinfo{person}{Jun Yan}, {and}
  \bibinfo{person}{Jian Zhang}.} \bibinfo{year}{2022}\natexlab{}.
\newblock \showarticletitle{ICCBot: Fragment-Aware and Context-Sensitive ICC
  Resolution for {Android} Applications}. In \bibinfo{booktitle}{\emph{The 44th
  International Conference on Software Engineering, {ICSE} 2022 (Tool Track)}}.
\newblock


\bibitem[\protect\citeauthoryear{Yang, Yan, Wu, Wang, and Rountev}{Yang
  et~al\mbox{.}}{2015a}]%
        {DBLP:conf/icse/YangYWWR15}
\bibfield{author}{\bibinfo{person}{Shengqian Yang}, \bibinfo{person}{Dacong
  Yan}, \bibinfo{person}{Haowei Wu}, \bibinfo{person}{Yan Wang}, {and}
  \bibinfo{person}{Atanas Rountev}.} \bibinfo{year}{2015}\natexlab{a}.
\newblock \showarticletitle{Static Control-Flow Analysis of User-Driven
  Callbacks in Android Applications}. In \bibinfo{booktitle}{\emph{37th
  {IEEE/ACM} International Conference on Software Engineering, {ICSE} 2015,
  Florence, Italy, May 16-24, 2015, Volume 1}}. \bibinfo{publisher}{{IEEE}
  Computer Society}, \bibinfo{pages}{89--99}.
\newblock


\bibitem[\protect\citeauthoryear{Yang, Zhang, Wu, Wang, Yan, and Rountev}{Yang
  et~al\mbox{.}}{2015b}]%
        {DBLP:conf/kbse/YangZWWYR15}
\bibfield{author}{\bibinfo{person}{Shengqian Yang}, \bibinfo{person}{Hailong
  Zhang}, \bibinfo{person}{Haowei Wu}, \bibinfo{person}{Yan Wang},
  \bibinfo{person}{Dacong Yan}, {and} \bibinfo{person}{Atanas Rountev}.}
  \bibinfo{year}{2015}\natexlab{b}.
\newblock \showarticletitle{Static Window Transition Graphs for {Android}}. In
  \bibinfo{booktitle}{\emph{{ASE} 2015}}. \bibinfo{pages}{658--668}.
\newblock


\bibitem[\protect\citeauthoryear{Yang and Yang}{Yang and Yang}{2012}]%
        {Leakminer}
\bibfield{author}{\bibinfo{person}{Zhemin Yang} {and} \bibinfo{person}{Min
  Yang}.} \bibinfo{year}{2012}\natexlab{}.
\newblock \showarticletitle{LeakMiner: Detect Information Leakage on Android
  with Static Taint Analysis}. In \bibinfo{booktitle}{\emph{2012 Third World
  Congress on Software Engineering}}. \bibinfo{pages}{101--104}.
\newblock
\urldef\tempurl%
\url{https://doi.org/10.1109/WCSE.2012.26}
\showDOI{\tempurl}


\bibitem[\protect\citeauthoryear{Zhang, Tian, and Duan}{Zhang
  et~al\mbox{.}}{2021a}]%
        {DBLP:journals/compsec/ZhangTD21}
\bibfield{author}{\bibinfo{person}{Jie Zhang}, \bibinfo{person}{Cong Tian},
  {and} \bibinfo{person}{Zhenhua Duan}.} \bibinfo{year}{2021}\natexlab{a}.
\newblock \showarticletitle{An efficient approach for taint analysis of android
  applications}.
\newblock \bibinfo{journal}{\emph{Comput. Secur.}}  \bibinfo{volume}{104}
  (\bibinfo{year}{2021}), \bibinfo{pages}{102161}.
\newblock
\urldef\tempurl%
\url{https://doi.org/10.1016/j.cose.2020.102161}
\showURL{%
\tempurl}


\bibitem[\protect\citeauthoryear{Zhang, Tian, Duan, and Zhao}{Zhang
  et~al\mbox{.}}{2021b}]%
        {DBLP:journals/tr/ZhangTDZ21}
\bibfield{author}{\bibinfo{person}{Jie Zhang}, \bibinfo{person}{Cong Tian},
  \bibinfo{person}{Zhenhua Duan}, {and} \bibinfo{person}{Liang Zhao}.}
  \bibinfo{year}{2021}\natexlab{b}.
\newblock \showarticletitle{RTPDroid: Detecting Implicitly Malicious Behaviors
  Under Runtime Permission Model}.
\newblock \bibinfo{journal}{\emph{{IEEE} Trans. Reliab.}} \bibinfo{volume}{70},
  \bibinfo{number}{3} (\bibinfo{year}{2021}), \bibinfo{pages}{1295--1308}.
\newblock


\bibitem[\protect\citeauthoryear{Zhang and Yin}{Zhang and Yin}{2014}]%
        {DBLP:conf/ndss/ZhangY14}
\bibfield{author}{\bibinfo{person}{Mu Zhang} {and} \bibinfo{person}{Heng Yin}.}
  \bibinfo{year}{2014}\natexlab{}.
\newblock \showarticletitle{AppSealer: Automatic Generation of
  Vulnerability-Specific Patches for Preventing Component Hijacking Attacks in
  Android Applications}. In \bibinfo{booktitle}{\emph{21st Annual Network and
  Distributed System Security Symposium, {NDSS} 2014, San Diego, California,
  USA, February 23-26, 2014}}. \bibinfo{publisher}{The Internet Society}.
\newblock


\end{thebibliography}
